\documentclass[acmsmall]{acmart}
\usepackage[linesnumbered,ruled,vlined]{algorithm2e}

\usepackage{multirow}
\usepackage{float}
\usepackage{subfig}
\usepackage{array}
\newcolumntype{H}{>{\setbox0=\hbox\bgroup}c<{\egroup}@{}}

\AtBeginDocument{%
  \providecommand\BibTeX{{%
    \normalfont B\kern-0.5em{\scshape i\kern-0.25em b}\kern-0.8em\TeX}}}

\setcopyright{acmcopyright}
\copyrightyear{2021}
\acmYear{2021}
\acmDOI{10.1145/1122445.1122456}

\acmJournal{JACM}
\acmVolume{37}
\acmNumber{4}
\acmArticle{111}
\acmMonth{8}

\begin{document}

\title{ReFRS: Resource-efficient Federated Recommender System for Dynamic and Diversified User Preferences}

\author{Mubashir Imran}
\email{m.imran@uq.net.au}
\orcid{0000-0003-4721-499X}
\affiliation{%
  \institution{The University of Queensland}
  \streetaddress{St Lucia}
  \city{Brisbane}
  \state{QLD}
  \country{Australia}
  \postcode{4072}
}
\author{Hongzhi Yin*}\thanks{*Corresponding author.}
\email{h.yin1@uq.edu.au}
\affiliation{%
  \institution{The University of Queensland}
  \streetaddress{St Lucia}
  \city{Brisbane}
  \state{QLD}
  \country{Australia}
  \postcode{4072}
}
\author{Tong Chen}
\email{tong.chen@uq.edu.au}
\affiliation{%
  \institution{The University of Queensland}
  \streetaddress{St Lucia}
  \city{Brisbane}
  \state{QLD}
  \country{Australia}
  \postcode{4072}
}

\author{Nguyen Quoc Viet Hung}
\affiliation{%
  \institution{Griffith University}
  \city{Gold Coast}
  \country{Australia}}
\email{henry.nguyen@griffith.edu.au}

\author{Alexander Zhou}
\affiliation{%
  \institution{Hong Kong University of Science and Technology}
  \city{Hong Kong}
  \country{China}
}
\email{atzhou@cse.ust.hk}

\author{Kai Zheng}
\affiliation{%
 \institution{University of Electronic Science and Technology of China}
 \city{Chengdu}
 \state{Sichuan}
 \country{China}}

\renewcommand{\shortauthors}{Mubashir, et al.}


\begin{abstract}
  Owing to its nature of scalability and privacy by design, federated learning (FL) has received increasing interest in decentralized deep learning. FL has also facilitated recent research on upscaling and privatizing personalized recommendation services, using on-device data to learn recommender models locally. These models are then aggregated globally to obtain a more performant model, while maintaining data privacy. Typically, federated recommender systems (FRSs) do not take into account the lack of resources and data availability at the end-devices. In addition, they assume that the interaction data between users and items is i.i.d. and stationary across end-devices (i.e., users), and that all local recommender models can be directly averaged without considering the user's behavioral diversity. However, in real scenarios, recommendations have to be made on end-devices with sparse interaction data and limited resources. Furthermore, users' preferences are heterogeneous and they frequently visit new items. This makes their personal preferences highly skewed, and the straightforwardly aggregated model is thus ill-posed for such non-i.i.d. data. In this paper, we propose \textit{Resource Efficient Federated Recommender System} (ReFRS) to enable decentralized recommendation with dynamic and diversified user preferences. On the device side, ReFRS consists of a lightweight self-supervised local model built upon the variational autoencoder for learning a user's temporal preference from a sequence of interacted items. On the server side, ReFRS utilizes a scalable semantic sampler to adaptively perform model aggregation within each identified cluster of similar users. The clustering module operates in an asynchronous and dynamic manner to support efficient global model update and cope with shifting user interests. As a result, ReFRS achieves superior performance in terms of both accuracy and scalability, as demonstrated by comparative experiments on real datasets.
\end{abstract}



\maketitle

\section{Introduction}\label{intro}
In the era of big data, recommender systems (RSs) play a pivotal role in handling information overload. Users of online services, such as e-commerce and video streaming, demand that  relevant items to be recommended upfront. In order to provide each user with personalized recommendations, users' personal information and behavioral footprints are collected and then hosted at a central server to facilitate user preference modelling. Typically, such data consists of user ID, age, gender, demographics, likes, dislikes and other interactions with items. As those attributes are highly personal and even sensitive, most users tend to be reluctant to share such sensitive information with their service provider \cite{leon2013matters}. Furthermore, laws such as the General Data Protection Regulation (GDPR) \cite{voigt2017eu} prohibit service providers from accessing unconsented user data.

Consequently, handling a massive volume of data and promising each user on data privacy introduces a dilemma for the current centralized RSs \cite{yang2020federated}, as users cannot enjoy accurate personalized recommendation without handing over their data to the server for analysis. In centralized systems, this trade-off between privacy and personalization \cite{awad2006personalization} is ineluctable, as they require both users' personal information as well as large-scale user-item interactions in order to facilitate training.
Additionally, as user-item interaction datasets easily exceed billion-scale, training RS on a central server raises scalability and efficiency issues, especially for state-of-the-art models that are commonly deep learning-based.

Recent advances in computational hardware of edge devices have given rise to the possibility of distributed machine learning paradigms, where a typical representative is federated learning (FL). In FL, user models are trained collaboratively on their local devices by exchanging only the local model parameters, without disclosing any raw user data to the server. The submitted local models are aggregated into a global model on the server side \cite{karimireddy2020scaffold,konevcny2016federated,lin2020meta}, and this global model is then re-distributed to the edge devices. Owing to FL's decentralized nature, it offers two advantages. First, FL relieves the burden of computational dependencies on servers and second, it ensures user privacy by design because the data is fully retained by the users.

Given the high privacy sensitivity and resource-intensive characteristics of recommendation services, FL has attracted an increasing amount of research attention in the field of RSs \cite{ammad2019federated,9162459,chen2018federated,niu2020billion}. Current federated recommender systems (FRSs) adopt different collaborative filtering methods as their local recommenders, and perform FL via variants of the model averaging algorithm \cite{mcmahan2017communication}, in order to obtain an expressive global recommender. The FRSs \cite{ammad2019federated,niu2020billion,9162459,jia2021personalized,wang2021demystifying} train the local models with the users' on-device data and collaborate with the federated server (FS) to iteratively improve the local model with the global context.
Despite the decentralization and privacy guarantees provided, in what follows, we point out several shortcomings of existing FRSs that remain unsolved and inevitably hinder their usability in practice.

Currently, FL-based architectures for RS \cite{wang2021fast,ali2021federated,ammad2019federated,niu2020billion,9162459,jia2021personalized,wang2021demystifying} are primarily focused on generating an accurate recommender model on users' devices. 
However, they fail to consider the fact that end-devices are usually limited in computational power and memory. For example, in federated MF, a device must store all item embeddings and (at least) one user embedding to facilitate model training and inference. The embeddings become a major memory bottleneck due to the large number of possible items in a dataset \cite{chen2021learning,li2021lightweight}. The 128-dim embedding of a single item takes approximately 10.2KB, and the embedding of just 10000 items takes more than 100MB memory. Hence, storing embedded representations of more than a million items on a single device would be infeasible. On the other hand, the parameters of a very simple General Matrix Factorization (GMF) model with 21,009 parameters take approximately 1.352 MB, on 32-bit systems accepting 128-dim input. Secondly, processing models with large size becomes computationally expensive. On edge devices with limited resources, such models become impractical as the number of items and model complexity increases.

Additionally, current FRS frameworks typically consider a scenario in which complete user interaction history is already available on the end-device, neglecting sequential nature of user-item interactions \cite{chen2019air,sun2020go,zhao2020discovering}. In the real-world, most user devices have sparse data (especially those with new users) \cite{tam2017retaining}, which limits the ability of local and global models to capture users' changing preferences or demands. 
Lastly, each user demonstrates strictly non-i.i.d interactions, revealing preferences that vary across cultures, genders, age groups, demographics, and so on. As a result, a single global model may not be enough to accurately reflect the heterogeneity in user behavior distributions.
As shown in the evaluation section, the federated model experiences reduced performance due to an inability to cater to both data sparsity and diverse user behavior/preference distributions.

\subsubsection*{\textbf{Motivation.}}
Overall, existing FRS are challenged by 1) poor quality of learned models due to data sparsity, 2) weak ability to model dynamic preferences, 3) limited personalization due to traditional FL's single global aggregation of models, and 4) significant memory consumption.
In light of these challenges, we aim to develop an FRS that is lightweight, handles data sparsity, captures dynamics of user interests, and considers heterogeneity of users during the global update. These challenges motivate use to design a more practical paradigm for learning distributed recommender models. To capture the dynamic user interests, we learn temporal representations of a user's preferences via her/his sequential interactions with items. For the diverse personal preferences and the resulted non-i.i.d on-device data, we dynamically cluster and aggregate models from semantically similar users instead of all users. Since each device shares model parameters with the server, we group users based upon these parameters rather than sharing users' personal information. 

\subsubsection*{\textbf{Our Contribution.}} To this end, we propose \textit{Resource-efficient Federated Recommender System} (ReFRS), an FRS that can effectively manage dynamic and diversified user preferences, even when the edge device is resource-constrained and subject to sparse user interactions. The framework of ReFRS is comprised primarily of two components, namely a decentralized client module and a semantic server.

Based on the sequential on-device interactions generated by the user, the locally deployed client module learns dynamic user interests.
To avoid storing all item information on the edge device for recommendation, we use a sequential approach where interactions are processed in sessions. These sessions are kept small to fit into limited memory efficiently, and old sessions are swapped out to maximize memory utilization.
This module consists of three parts. An interaction sampler in the client module collects and organizes the user's interactions into chunks of stacked sequences. Each of these chunks is converted into a low-dimensional vector representation by an embedding model. The embedding model is a lightweight generative model based on a self-supervised variational autoencoder (VAE) \cite{imran2020decentralized}, which condenses continuous vectors into discrete vectors using vector quantization (VQ) \cite{DBLP:conf/acl/JinWSL20}. A hidden representation of each interaction sequence is extracted from the VQ layer as an embedded vector. It reflects the most distinct characteristics of user's temporal interactions and is fed as an input to the recommender model. The model parameters of the encoder layer are shared with the federated server, which enriches the local models with high-order information. It allows faster convergence of the local models and is communication efficient since we only have to share a single layer's parameters.
A major advantage of our client model is that it can effectively handle sparse data and convert it into a size-constrained discrete vector representation.

To overcome the non-i.i.d. nature of individual user data, we propose an efficient grouping scheme which only aggregates client models that are semantically similar. Rather than grouping clients by requesting and comparing all users' sensitive information like demographic data \cite{muhammad2020fedfast}, we utilize a neural approach to efficiently capture user affinity by simply looking at the model parameters submitted by each client. 
This is accomplished via a federated server, which is composed of a semantic sampler, a clustering unit, and an aggregator.
Semantic samplers contain a VAE, which encodes every user's model parameters into a compact vector representation, maintaining close affinity among semantically similar user models in the embedding space. A dynamic clustering mechanism is then used to group these low-level vector representations of clients' models. Based on the low-level representation of clients' model parameters, the clustering unit is able to generate discriminative and dynamic clusters. The aggregator then aggregates all the clients' models within each cluster and then distributes the cluster-wise global model to all the clients in the same cluster. This approach aggregates users with similar interests. All three components of the federated server are scalable and communicate asynchronously.
The dynamic nature of our clustering unit allows for the adaptive increase and decrease in the number of clusters based on the variance in the incoming user models. As the cluster formation is asynchronous, the communication bandwidth is not affected.

One of the advantages of FL is that no amount of user information or generated data is shared during the model federation. However, the shared model parameters are very vulnerable to membership inference attacks \cite{zhang2021membership,shokri2017membership} and attribute inference attacks  \cite{carlini2019secret,yeom2018privacy} on the server. This may lead to the leakage of sensitive or private information. For this reason, we adopt Differential Privacy (DP) \cite{wei2020federated} and Homomorphic Encryption (HP) \cite{jiang2021flashe} to showcase the strong compatibility of ReFRS with these privacy schemes.
Our main contributions are summarised as follows:
\begin{itemize}
\item We design a lightweight generative model that learns discrete item representations conditioned on the temporal context, and is able to fully capture the dynamic user preferences for sequential recommendation.
\item We propose an asynchronous and scalable model aggregation approach for the federated server, allowing for grouping semantically similar user models into clusters. The global models are aggregated cluster-wise to support quicker local model convergence and better personalized recommendation in the presence of non-i.i.d. data.
\item  We validate the advantageous effectiveness and scalability of ReFRS by conducting extensive experiments with real-world datasets of different sizes and state-of-the-art baselines.
\end{itemize}

\section{Related Work}\label{relatedwork}
\subsection{Recommender Systems (RS)}
A RS gathers information about the preferences of its users for a set of items e.g., movies, songs, books, jokes, gadgets, applications, websites, etc. The information can be collected either explicitly (typically from the ratings of users) or implicitly (typically by monitoring the users' actions, such as the songs they hear, the applications they download, or the websites they visit). RS may also use demographic information such as age, nationality, and gender, as well as social information, like followers, followed, tweets, and posts, that is commonly used in Web 2.0, to predict users' interests. 
As a general rule, recommender systems are programs that attempt to predict an individual or business' interest in an item based on information about the item, the user, and their interactions with the item.

In the mid-1990s, researchers began to focus on recommendation problems based on explicit ratings, and recommender systems emerged as an independent research area \cite{adomavicius2005toward,goldberg1992using}. The most common recommendation techniques are collaborative filtering (CF) \cite{schafer2007collaborative}, the content-based approach (CB) \cite{pazzani2007content} and knowledge-based approach (KB) \cite{burke2000knowledge}. In addition to its strengths and weaknesses, each recommendation method has advantages and disadvantages; e.g., CF suffers from sparseness, scalability, and cold-start problems \cite{adomavicius2005toward,schafer2007collaborative}, while CB offers overspecialized recommendations \cite{adomavicius2005toward,goldberg1992using}. Many advanced recommendation approaches have been proposed to solve these problems, including social network-based recommender systems \cite{he2010social}, fuzzy recommender systems \cite{zhang2013hybrid}, context-aware recommender systems \cite{adomavicius2011context}, group recommender systems \cite{masthoff2011group}, session based recommender systems \cite{qiu2020exploiting} and sequential recommender systems \cite{tang2018personalized}.

\subsection{Sequential Recommender Systems (SRS)} 
In Sequential Recommender Systems (SRS), each user is associated with a number of interaction sequences with some items. Its goal is to recommend each user the  most probable next item by considering that user's general tastes as well as short-term intention \cite{tang2018personalized}.
Most of the early research on SRS adopts the Markov Chain (MC) \cite{mobasher2002using,rendle2010factorizing,shani2005mdp} method to model sequential behavior by estimating an item-item transition probability, and predicting the next item. The introduction of neural networks paved way for learning based RS, with the adoption of Gated Recurrent Units (GRU) \cite{hidasi2015session}  to the session-based recommendation. Modifications of GRU based methods soon followed, by adopting  pair-wise loss functions \cite{hidasi2016parallel}, memory networks \cite{huang2019taxonomy,huang2018improving}, hierarchical structures \cite{quadrana2017personalizing}, copy mechanism \cite{ren2019repeatnet} and reinforcement learning \cite{zhou2020s3}, etc. 

Recently generative model have generated increased interest in SRS, that captures user behavior and time dependencies in those preferences. One such approach is developed by SVAE \cite{sachdeva2019sequential}, which  utilizes variational autoencoders for modeling a user's preferences by incorporating latent variables and temporal dependencies. It models latent dependencies by using recurrent neural networks, before feeding them into the VAE model for prediction. Another popular approch is 
To generate high-quality latent variables, ACVAE \cite{xie2021adversarial} adopts the Adversarial Variational Bayes (AVB) \cite{mescheder2017adversarial} framework. Then, it applies a contrastive loss \cite{inoue2020semi} to compare the latent variables between different users.
Contrary to the use of multi-user single models in these proposed models, ReFRS follows a federated setting, where each user only has access to their own data. As part of the embedding model, ReFRS takes in a window of recently interacted items and learns their temporal representation using a generative model. The vector quantization (VQ) layer and the multi-head attention between the encoder latents and the quantized latent help preserve the salient temporal features and interaction dependencies.

\subsection{Federated Learning (FL)}
FL is an online architecture where under the coordination of a central aggregator, multiple clients can collaborate to solve machine learning (ML). FL allows the training to be carried out on a local device, in a decentralized fashion, to ensure the data privacy of each device \cite{zhang2021survey}. The main contribution of FL is preserving user privacy while allowing multiple users to collaborate.  FL architecture can be divided into three categories, horizontal FL, vertical FL and federated transfer
learning \cite{yang2019federated}. Horizontal FL is suitable for a scenario where the items that are interacted by the users overlap but not the users themselves \cite{bonawitz2017practical,smith2017federated}. In Vertical FL, the items being interacted with only overlap a little (if any). However, the users (or user interactions) have a high degree of overlap \cite{cheng2021secureboost}. Finally, in federated transfer learning, both the users and the user-item interactions rarely overlap. In this case, rather than segmenting the data, we adopt transfer learning for the lack of data or tags \cite{liu2020secure}. Model aggregation is one of the  main components in FL, which trains the global model by summarizing the model parameters from all participating clients \cite{zhang2021survey}.

\subsection{Federated Recommender System(FRS)}
RSs are primarily data-driven systems, where the common case is that the more data it uses, the better it performs. However, due to privacy and security constraints, directly sharing user data between parties is undesired \cite{tan2020federated}. FRS builds up a stronger recommender model without compromising user privacy or data security. Recently, Ammad et al. \cite{ammad2019federated} introduced the first FL-based collaborating filtering (CF) method, that utilizes stochastic gradient descent (SGD) to update the local model and the FedAvg mechanism is adopted to update the global model. Improving on data security and potential leakages in \cite{ammad2019federated}, \cite{9162459} introduces FedMF, which enhances the distributed matrix factorization (MF) framework  to the FL setting and apply homomorphic encryption on the gradient information before sharing the information. FedMeta \cite{chen2018federated} is a federated meta-learning framework, where a parameterized algorithm is shared as compared to a global model using FedAvg, adopted in a typical FL setting. SFSL \cite{niu2020billion} observe that clients interact with a subspace of items rather than complete feature space. To cater effective utilization of bandwidth and ensure model anonymity, they design a secure federated sub-model learning mechanism coupled with a private set union protocol. FedFast \cite{muhammad2020fedfast} accelerates federated recommendation tasks by applying an active aggregation method that propagates the updated models to similar clients. However, FedFast \cite{muhammad2020fedfast} relies on user grouping based upon profile similarity. This similarity is computed by sharing user's personal demographic information with the central server, leaving users vulnerable to identity leakage \cite{kendzierskyj2019transparency}. 

Some clustering based FL approaches have been proposed that support users' behavioral heterogeneity without sharing/leaking any user data. Iterative Federated Cluster Algorithm (IFCA) \cite{ghosh2020efficient} attempts to minimize the loss function of each FL client while also assigning the client to a cluster. Two variants are proposed for model aggregation in IFCA, i.e., model and gradient averaging. IFCA applies random initialization and multiple restarts to cluster clients and reach optimal results. However, IFCA  estimates each client's cluster by iteratively finding the best performing model parameters on the local device. Consequently, this will incur additional communication and computational costs on resource-constrained user devices like cellphones or tablets. A more recent clustering-based FL approach, DistFL \cite{liu2021distfl}, extracts and compares distribution knowledge from uploaded models to perform clustering based on the data generated from uploaded models, but its practicality is also constrained by the high communication overhead. Furthermore, in IFCA, each user device participating in federated clustering has to be synchronously on the same training step. In the real world, this is infeasible, especially in recommendation systems where user devices are constantly connecting and disconnecting from the server.

\section{PRELIMINARIES}\label{prem}
In this section, for the ease of understanding, we formally define some key technical terms, followed by problem formulation and system overview. The notations adopted to describe key technical elements are listed in \autoref{notations}.

\begin{table}[h]
\caption{Notations used throughout the paper.}
\label{notations}
\begin{tabular}{c|l}

\hline
Notation          & \multicolumn{1}{c}{Description}                                             \\ \hline
$\mathcal{U}$     & Set of users                                                                \\
$n$               & \begin{tabular}[c]{@{}l@{}}Number of users in  $\mathcal{U}$\end{tabular} \\
$\mathcal{I}$     & Set of items                                                                \\
$m$               & Number of users in $\mathcal{I}$                                            \\
$F_{i}$           & Feature set of item $i$                                                     \\
$d$               & Number of dimensions                                                        \\
$s_j$             & Sequence $j$, representing number of items interacted with within a window  \\
$w$               & Consective number of items                                                  \\
$\mathcal{S}_{u}$ & Session, contains a collection of consective sequences                      \\
$\mathbf{x}_s$             & Input feature vector, $x_s = [F_{i_1};F_{i_2};..;F_{i_w}]$ of items in $s$  \\
$\mathbf{z}$               & Latent representation of $x$                                                \\
$\mathbf{e}$               & Embedding vector of an interaction                                          \\ \hline
\end{tabular}
\end{table}

\noindent \textbf{Term 1:} \textit{Edge Device} is the digital hardware used by the user for streaming and other services. It could be a desktop computer, laptop, tablet, etc. We also refer to the user's device as a client in this work.

\noindent \textbf{Term 2:} \textit{Global Epoch} represents a full iteration in which all the participating clients have interacted with the global server.

\noindent \textbf{Term 3:} \textit{Asynchronous Clustering} represents the computation of client clusters based on their local models' parameters, which happens when the uploaded local models are being aggregated into the global model and then transmitted back to local devices.

\noindent \textbf{Term 4:} \textit{Aggregated Global Models} are the aggregation of all client models within \textit{each cluster}. The total number of aggregated models are determined by the number of clusters, which changes automatically with each asynchronous clustering commit. 

\begin{figure}[ht]
	\centering
	\includegraphics[width=\linewidth]{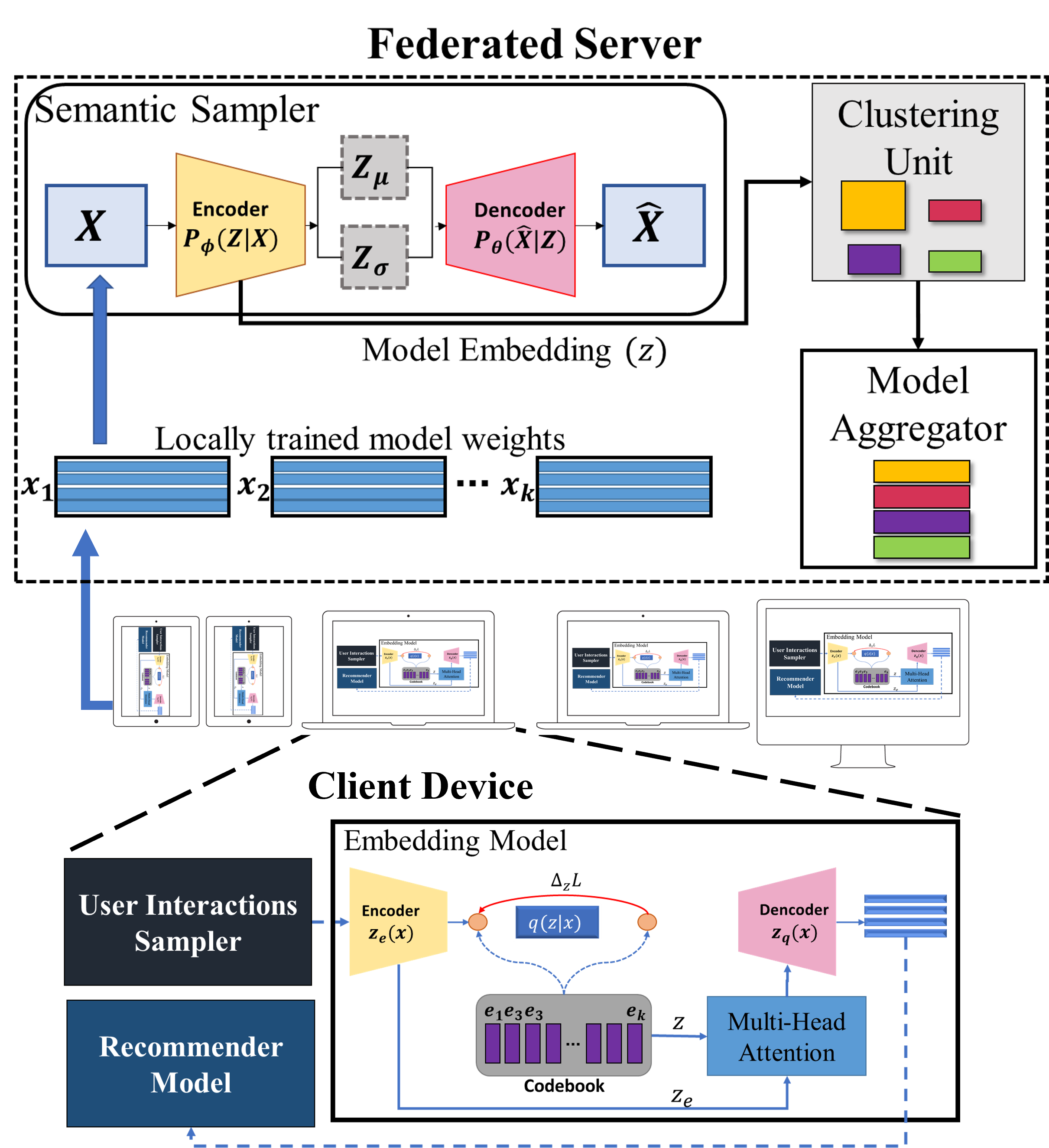}
	\caption{ A central federated server, responsible for 1) identifying similarity among clients by learning the structure of their model parameters, 2) clustering client models based upon parameter similarity and 3) aggregating and distributing the clustered model parameters. Local recommender model has three parts. The first part consists of interaction sampler that stacks sequences of interactions within a given window, the second part consists of a self-supervised generative model that learns discrete latent representations of item interaction and the third part supports a recommender model that uses the item embeddings to forecast next item.}
	\label{fig:model}
\end{figure}

\subsection{Problem Formulation}
Let a set of $n$ users $\mathcal{U}$ interact independently with a set of $m$ items $\mathcal{I}$, and each $i\in \mathcal{I}$ has features $\mathbf{F}_{i} \in \mathbb{R}^d$, where $d$ is the number of dimensions. Each user $u\in \mathcal{U}$ accumulates various sessions $\mathcal{S}_{u} =\{s_{1}, s_{2}, ...\}$ on her/his local machine over time, where a session $s=\{i_1,i_2,...i_w\}$ is a sequence of $w$ interacted items in a short time period. Note that we use numbered subscripts for items mainly to indicate their orders in a sequence. Our sequential recommendation task is session-based, where the model predicts the next item $i_{w+1}$ for every given $s \in \mathcal{S}_u$. Due to multi-faceted user preferences, the distribution of $\mathcal{S}_u$ is highly skewed across users. Also, compared with the centralized setting where all users' data is available, in FRS, each user has a highly limited amount of interactions for learning her/his preference. To fully capture such data heterogeneity, we perform asynchronous clustering to help generate aggregated global models, so as to achieve better personalized recommendation accuracy in an effective and efficient way. 

\subsection{System Overview}

ReFRS follows the standard FL client-server architecture, as illustrated in Section \ref{fig:model}. Each client corresponds to a personal device that holds a single user's interaction data and trains a lightweight recommender model. A client hosts an interaction sampler, an embedding model and a recommender model. The server identifies client groups based on their model parameters and keeps a copy of the aggregated recommender model for each group. Overall the server is responsible for: 1) identifying, grouping and asynchronously updating client groups based on their parameter similarity; 2) aggregating and disseminating model parameters within each user cluster. The detailed designs of both the client and server are presented in Section \ref{client_model} and Section \ref{server_model}. 

\section{Client-side Dynamic Preference Modelling}\label{client_model}
We start the description of ReFRS with the locally deployed model on end-devices, which learns each user's dynamic preferences via sequential interactions $s \in \mathcal{S}_u$. The client module consists of three parts: 1) user interaction sampler, 2) an embedding model and 3) a recommender model.
We assume that each client device holds a single user's data. The objective of user preference embedding is to aggregate historical user-item interactions in order to determine the user's preferences. The ReFRS client comprises a self-supervised embedding module that encodes these preferences i.e., the relation between the items. The embedded vectors computed by the self-supervised embedding module are then fed into a recommender model that predicts the next item. 

\subsection{User Interactions Sampler}
\label{section:seq}
In the federated setting, each client holds only its own user's interaction sequences $\mathcal{S}_u$. These sequences are first broken down into smaller sessions in order to train the user model. To simulate sequential behavior, these sessions are fed to the embedding model as batches.
Since the same item may recur over time in a sequential system, it is important to take note of how this item co-occurs with its neighbors over time \cite{chorney2010time}. This results in different representations of an item in different sessions, reflecting changes in the user's behavior. For this reason we define an interaction window $s = \{i_1,i_2,...i_{|w|}\}$, which represents the interaction history of an item $i_{w+1}$. Taking into account that  $s=\{i_1,i_2,..,i_w\}$ consists of $w$ items and leads to the $(w+1)^{th}$ item, we model the input to our embedding model using sequential feature vectors $\mathbf{x}_w = [\mathbf{F}_{i_1};\mathbf{F}_{i_2};..;\mathbf{F}_{i_w}]$ of items in $s$, used to estimate the latent representation $\mathbf{z}_{w+1}\in  \mathbb{R}^d$ of item $i_{w+1}$. 

\begin{figure}[ht]
	\centering
	\includegraphics[width=\linewidth]{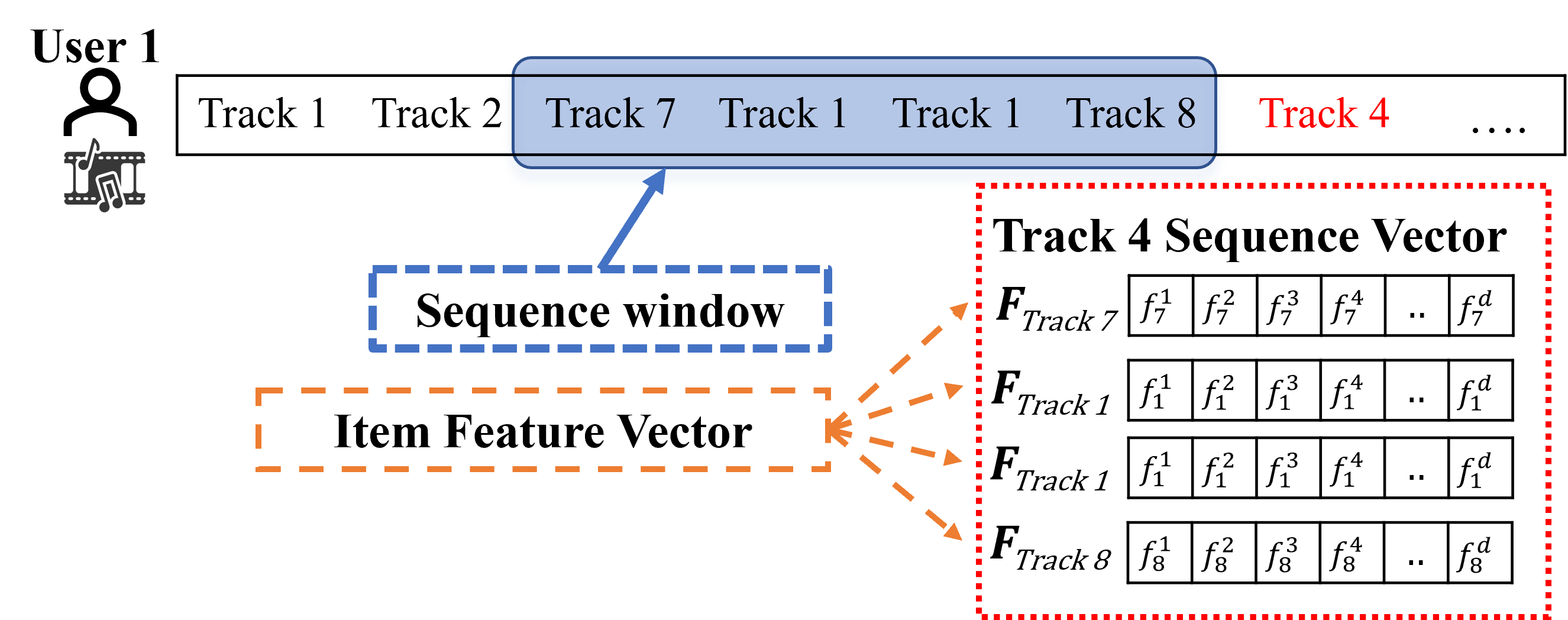}
	\caption{Top bar shows \textit{User 1}\textquoteright s interaction with multiple sound tracks over a period of time. The blue selection represents interaction window for \textit{Track 4}. At the bottom we have feature vector of \textit{Track 4}'s interaction window.}
	\label{fig:sequence}
\end{figure}

An example of aforementioned sequence sampling is given in \autoref{fig:sequence}.
\textit{User 1} interacts with multiple music tracks over a period of time. In order to get the user interaction behavior for \textit{Track 4} at time interval $t_7$, we define $w$ (of size 4 for the purpose of this example) and stack the feature vectors of tracks falling under the window $w$. Note that the generation of \textit{Track 4}'s embedding is conditioned on $\mathbf{x}_s$ (i.e., stacked features of \textit{Tracks 7, 1, 1, and 8} in this case).
The embedding of items in such sampling is influenced by their sequential context and temporal positioning. As such, the same item will have a different temporal embedding when different temporal contexts are given. The main advantage of our approach is that it captures the user's interest in an item $i_a$ at time $t$ when it is followed by sampled items $\{(i_b,i_c,i_d,..),(i_x,i_y,i_z,..),...\}$, grouped temporally. This change of user interest and variation in grouping persist in SRS setting, so it is beneficial to encode each item based on the latest context, when predicting the next item. In our work we define the best $w = 16$ after experimentation (using $w = 4,8,16$ and $32$). In the first session (i.e. batch), 16 interactions are sampled as $s_{i_1} = (i_{-15},i_{-14},...i_{0})$ for $i_1$, $s_{i_1} = (i_{-14},i_{-13},...i_{1})$ for $i_2$, and so on.

\subsection{Embedding Model}
\label{embedding_model}

Given a session, the embedding model learns informative item representations by incorporating the dynamic context carried by other recently interacted items.
In other words, we need to learn a compact vector representation (i.e., embedding) $\mathbf{z}$ of item $i_n$ from the interaction window $w$. The window groups semantically similar items together to capture the temporal and contextual relationships between the interacted items.
As a result of this representation, item-sequence relationships should be effectively captured and items that are contextually and temporally similar should be grouped together. Since an item can be repeatedly interacted with over time, it is important to fully incorporate such dynamic context into an item's embedding. Hence, in ReFRS, instead of using a fixed embedding for every item, each item's embedding is dynamically conditioned on its predecessors within the same session. 
In order to meet these requirements, we propose to utilize the sequential variational autoencoders (VAE) \cite{shao2021controlvae} to capture the relationship among the input feature vector. The VAE will try to capture the dependencies among the temporally stacked item sequence (as described in Section \ref{section:seq}), by enforcing a probabilistic prior to the latent space of the autoencoder \cite{zhu2020s3vae}. That is, instead of mapping the session $s$ to a point in the embedding space, VAE maps it to a normal distribution, making the representation $\mathbf{z}$ learned from similar sources close, compact and smooth to interpolate between. VAE uses Kullback-Leibler (KL) divergence to measure the information discrepancy between the approximated true posterior $p(\mathbf{x}|z)$ of the latent variables and the variational posterior $q(\mathbf{z}|s)$:
\begin{equation}
\label{eq:embedding_VAE}
	 \textbf{KL}\left(q\left(\mathbf{z}_{w+1}|\mathbf{x}_s\right)||p\left(\mathbf{z}_{w+1}\right)\right)-\textbf{E}_{\mathbf{z}_{w+1} \sim q\left(\mathbf{z}_{w+1}|\mathbf{x}_s\right)}\left[\log \left(p\left(\mathbf{z}_s|\mathbf{z}_{w+1}\right)\right)\right]
\end{equation}
where $q(\mathbf{z}_{w+1}|\mathbf{x}_s)$ denotes posteriors approximated by the neural network encoder, and $p(\mathbf{x}_s|\mathbf{z}_{w+1})$ is the reconstruction approximated by the decoder, which is also a neural network.

However, in federated settings where there is only one user's data (a small amount of data that grows slowly), the training objective of VAE (i.e., KL divergence) quickly reaches local maxima \cite{lucas2019don}. As a result, the continuous latent representation learned by VAE suffers from issues like high variance and posterior collapse \cite{dai2020usual}. Furthermore, the continuous vector representation produced by VAE requires a large embedding representation to accurately represent the feature vector. However, the limited memory of the end-device, makes storing large embedding vectors counterproductive.

\begin{figure}[ht]
	\centering
	\includegraphics[width=\linewidth]{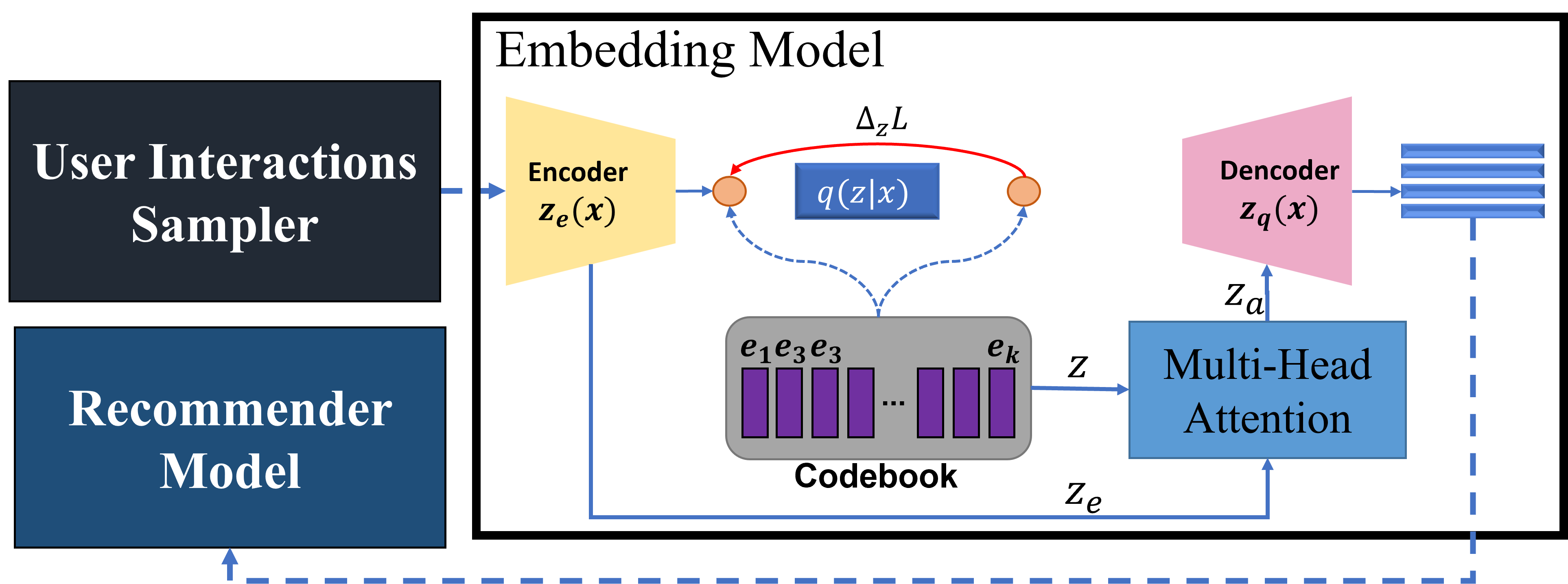}
	\caption{The client embedding model consists of an encoder model having 1) Encoder (consisting of 3 stacked CNN layers), 2) Vector Quantization Layer (having code book for mapping continuous to discrete latents and vice versa), and 3) Decoder (reconstructs the original sequence).}
	\label{fig:client_model}
\end{figure}

We propose to use the vector quantization (VQ) \cite{DBLP:conf/acl/JinWSL20} technique to discretize the latent variable space into numerous sub-spaces, and further adopt self-supervision to drive the learning of discrete latent variables.
This technique effectively projects each interaction window $w$ in a separate sub-space. VQ transforms continuous vectors into finite sets of "code" vectors. In essence, it works the same way as k-nearest neighbors (KNN), in that a sample is mapped to a centroid code vector with the minimum Euclidean distance.
The resultant discrete latent representations are also more constricted and space-efficient. 

As shown in \autoref{fig:client_model}, an encoder layer converts the input $\mathbf{x}$ into latents $\mathbf{z}_e$ and the VQ layer then convert $\mathbf{z}_e$ into discrete latents $\mathbf{z}$ by calculating nearest neighbour look-up. Before feeding $\mathbf{z}$ to the decoder, we perform multi-head attention \cite{vaswani2017attention} on $\mathbf{z}_e$ and $\mathbf{z}$, yielding $\mathbf{z}_a$, which is converted back in $\mathbf{x}$ by the decoder. As a result, the model can jointly attend to information from different representation from the latent sub-spaces at different positions. 
\begin{equation}
	\label{eq_attn}
	\begin{aligned}
	MultiHead&(\mathbf{z}, \mathbf{z}_e, \mathbf{z}_e ) = Concat(head_1, ..., head_h)\mathbf{W}^0\\
	& head_i = Attention({\mathbf{z}}\mathbf{W}^{\mathbf{z}}_i,{\mathbf{z}_e}\mathbf{W}^{\mathbf{z}_e}_i,{\mathbf{z}_e}\mathbf{W}^{\mathbf{z}_e}_i)
    \end{aligned}
\end{equation}

The discrete latents, fed into the attention network are the sampled distributions indexed by an embedding table \cite{van2017neural}. Overall the embedding model can be viewed as a VAE where $\log p(\mathbf{x}_s)$ is bounded by the evidence lower bound (ELBO) \cite{yang2017understanding}, as given in Eq. (\ref{eq_ELBO}):
\begin{equation}
	\label{eq_ELBO}
	q(\mathbf{z}=k|s) = \begin{cases}
		1 & \text{if}\,\, k=argmin_j||\mathbf{z}_{encode}(\mathbf{x}_s)-\mathbf{e}_j||_2\\
		0 & \text{otherwise}
		\end{cases}
\end{equation}
where $\mathbf{z}_{encode}(\mathbf{x}_s)$ denotes the output from the encoder given by $\mathbf{x}_s$, which is discretized by mapping to the nearest element of embedding $\mathbf{e}_j$, with $j$ given as:
\begin{equation}
	\label{eq_discretized}
	j = argmin_{j\in \{1,2,...,J\}}||\mathbf{z}_{encode}(\mathbf{x}_s)-\mathbf{e}_j||_2
\end{equation}
where each embedding vector $\mathbf{e}_j$ is drawn from the VQ's embedding table (a.k.a. codebook) consisting of $J$ discrete values. The objective function of our embedding model is then to minimize the flipped ELBO \cite{DBLP:conf/acl/JinWSL20} assuming degenerate approximate posteriors: 
\begin{equation}
	\label{eq_VQ_loss}
\begin{split}
	\mathbf{L}_{emb} &= -\log p(\mathbf{x}_s|\mathbf{z}_{encode}(\mathbf{x}_s))+||sg(\mathbf{z}_{encode}(s))-\mathbf{e}_j||^2_2 \\
	&+ \beta||\mathbf{z}_{encode}(s)-\mathbf{e}_j||^2_2.
\end{split}
\end{equation}

In Eq.(\ref{eq_VQ_loss}), the first term is the reconstruction loss from the decoder. While both the second and third terms use $l_2$ error to move $\mathbf{e}$ towards $\mathbf{z}_{encode}(\mathbf{x})$, in the second term we adopt the stop-gradient operator $sg(\cdot)$ \cite{DBLP:conf/acl/JinWSL20} to encourage the encoder output to commit as much as possible to its closest codebook vector. $\beta$ is a hyperparameter controlling the effect of the third term. Note that, once we obtain a well-trained embedding model, its parameters will stay fixed and we use the learned encoder $\mathbf{z}_{encode}(\cdot)$ followed by the vector quantization described in Eq.(\ref{eq_discretized}) to directly generate the temporal embeddings for observed items. With embeddings generated for all items in $s$, we further use a sequential recommender model to capture the dynamic preferences of the user. Note that, every time an item is visited, its feature vector is obtained from the server and a unique embedding is calculated based on its temporal position. As a result, all item embeddings are private and specific to the device. 

\subsection{Recommender Model}
\label{Recommender_model}

Both models for item embedding and next-item recommendation are trained independently. We follow the commonly used recommendation strategy here, i.e., predicting the next most likely item to be interacted with. Given our primary contribution was obtaining meaningful temporal embeddings, we test its effectiveness by feeding learned embedding to state-of-the-art sequential recommender models. The sequence representation ${h}$ (where, ${h}_{w+1}=\{\mathbf{z}_1\oplus \mathbf{z}_2\oplus...\mathbf{z}_w\}$) and the item embedding matrix $\textbf{V}=\{\mathbf{z}_1,\mathbf{z}_2,...\}$ are used to compute the predictive score as: 
\begin{equation}
	\label{eq_req}
	\hat{y}=\Omega({\textbf{V}h})
\end{equation}
In our work we use GRU4Rec \cite{hidasi2016parallel} and SASRec \cite{kang2018self} since they are the most widely adopted SRSs. In addition the learning mechanism proposed by these frameworks i.e., GRU and the transformer, are widely adopted in different variations by other state-of-the-art SRSs.


\section{Server-side Adaptive Model Aggregation}\label{server_model}
\label{server_model}
The model presented in Section \ref{client_model} is locally deployed on each client and is subject to data sparsity issues. In order to better generalize the local models and better capture the interaction context while reducing the communication cost for model aggregation, we only share the parameters of the encoder (of client embedding model) with the server. In order to maintain users' privacy, the user information, interaction data, item embeddings and model hyperparameters for the decoder and the codebook are strictly not shared.
Since the embedding model essentially maps each user's interaction behavior into the target item's embedding, it benefits the federated server by helping to identify similar user patterns. However, as described earlier, notwithstanding data heterogeneity and user preference diversity, most FRSs \cite{ammad2019federated,niu2020billion,9162459,jia2021personalized,wang2021demystifying} simply assume that the parameters of all user models can be directly aggregated. To overcome the non-i.i.d. nature of individual user data, we propose an efficient grouping scheme that only aggregates client models that are semantically similar. Rather than grouping clients by requesting and comparing all users' sensitive information like demographic data \cite{muhammad2020fedfast}, we utilize a neural approach to efficiently capture user affinity by simply looking at the model parameters submitted by each client. We also enable the server to asynchronously update client groupings as the user interaction behavior changes over time. Our sever-side design has three parts, namely a semantic sampler, a clustering unit and a model aggregator, which we will describe below.

\subsection{Semantic Sampler}
\label{semantic_sampler}
In FRSs where the main objective is to provide users with suggested items based on their historical interactions with a collection of items, the issue of non-i.i.d. data arises. For example, music streaming clients may be interested in different categories of music due to personal interests, demographics, moods etc. Thus, leveraging the heterogeneity in user behavior is of utmost interest. FL facilitates aggregating models for learning user preferences while maintaining each user\textquoteright s privacy, however by straightforwardly treating interactions from all the users as i.i.d., it inevitably harms the personalization characteristics of a recommender model. 

In light of this, we tackle the dilemma of interaction scarcity at the user level and non-i.i.d. data at the server level by investigating communities consisting of similar users. Intuitively, it would be beneficial to perform model aggregation within each of the user groups, such that each client model can be enhanced using only the knowledge from users with similar preferences. Though similar users can be easily identified via their interaction records or personal attributes, it will void the privacy guarantees provided by FRSs. So, we propose a semantic sampler that utilizes the shared model parameters to generate user clusters. More precisely, the semantic sampler treats the encoder model parameters of each client as its respective features, and embeds them into a low-dimensional vector representation via a stand-alone VAE model. The input to this VAE are the shared encoder model parameters (obtained in Section \ref{embedding_model}) of each client, obtained from encoder layer.
By obtaining low-dimensional vectors withholding client-wise similarity information, we are able to avoid heavy computation cost of directly clustering the high-dimensional model parameters. 
 
Using VAE, latent variables of a client $u$ are drawn from a prior $p(\mathbf{e}_u)$, where $\mathbf{e}_u$ is the embedding vector of the client's shared model parameters $\mathbf{m}_u$. Here $\mathbf{m}_u$ are flattened encoder layer parameters of client $u$. VAE defines a joint probability distribution $p(\mathbf{m}_u|\mathbf{e}_u)$ over the input data and latent variables, with the goal of inferring the latent variables (i.e., client's model embedding) given observed data (i.e., the client's model parameters): 
\begin{equation}
	p(\mathbf{e}_u|\mathbf{m}_u) = \frac{p(\mathbf{m}_u|\mathbf{e}_u)}{p(\mathbf{e}_u)},
\end{equation}
where the optimization goal can be formulated analogously to Eq.(\ref{eq:embedding_VAE}) using Kullback-Leibler (KL) divergence:
\begin{equation}
	\!\!\mathbf{L}_{sem}= \textbf{KL}\left(q\left(\mathbf{e}_u|\mathbf{m}_u\right)||p\left(\mathbf{e}_u\right)\right)-\textbf{E}_{\mathbf{e}_u \sim q\left(\mathbf{e}_u|\mathbf{m}_u\right)}\left[\log \left(p\left(\mathbf{m}_u|\mathbf{e}_u\right)\right)\right].
\end{equation}

As such, the server is able to compute low-dimensional vector representations $\mathbf{z}_u$ using model parameters shared by clients. In addition to reduced dimensionality, the VAE also maps similar models closer in the embedding space.

\subsection{Clustering Unit}
\label{Asynchronous_Clustering}
The semantic sampler computes a low-dimensional vector representation of each client's model parameters, as soon as they arrive at the server side. Note that in Section \ref{semantic_sampler}, vector embeddings are computed in such a way that semantically similar models will have close representations in the vector space and dissimilar clients will be mapped far apart. Such placement of model parameters in the embedding space fulfils both clustering properties \cite{bateni2017affinity,hung2017computing}, i.e. intra-cluster maximization and inter-cluster minimization. Making use of $k$-Means \cite{Jin2010} algorithm to compute effective clusters, we  group together similar client models. However, directly applying $k$-Means clustering incurs the following overheads: 1) the optimal value of $k$ is unknown; 2) it needs to be recomputed every time a new client has to be added; and 3) it is inefficient to synchronously update for each incoming client model.

Our asynchronous clustering module effectively addresses these challenges via a dynamic setting. After the first global epoch, using a small pool of sampled initial clients, the server trains the VAE-based semantic sampler and also generates initial client embeddings. With these initial embeddings, we run the Elbow method \cite{syakur2018integration} that uses $k$-Means clustering on the dataset, and then compute an average silhouette score \cite{de2015recovering} for all the clusters. Subsequently, the server obtains the incoming client\textquoteright s model embeddings by feeding its model parameters to the trained semantic sampler. For the subsequent epochs, we compute the Euclidean distance between cluster centroids and the embedded vector to efficiently assign/update the clustering outcome. Since the computation of the optimal clustering coefficient $k$ and the assignment of an incoming client model to an existing group are performed asynchronously to the model aggregation, the communication bandwidth of the federated server is not affected.
This live computation of client\textquotesingle s cluster, given in Algorithm \ref{algo_2}, simultaneously supports model scalability and efficiency.

\begin{algorithm}	
	\caption{Client Clustering}
	\label{algo_2}
	\SetKwData{Left}{left}\SetKwData{This}{this}\SetKwData{Up}{up}
	\SetKwFor{ForAll}{for $\forall$}{do in Parallel}{end}
	\SetKwFunction{Union}{Union}\SetKwFunction{FindCompress}{FindCompress}
	\SetKwInOut{Input}{input}\SetKwInOut{Output}{output}
	\KwIn{Semantic sampler $VAE(\cdot)$, all clusters' centroids $\mathcal{V}$, a client model's parameters $\mathbf{m}_u$}
	\KwOut{\textit{Cluster ID $c_u$ for client $u$}}
	\BlankLine 

	 $\vartriangleright$ \textbf{Cluster Assignment During Global Epoch} \\
	$\mathbf{Model} \leftarrow VAE(\{\mathbf{m}_1,\mathbf{m}_2,...\mathbf{m}_u\})$\\
	$\mathbf{e}_u \leftarrow \mathbf{Model}[(\mathbf{m}_u)]$\\

	$\{c_{1},c_{2},... c_{k}\}\leftarrow Elbow({\mathbf{e}_1,\mathbf{e}_2,...\mathbf{e}_u})$ \tcp*{Compute optimal number of Clusters}
	
	
	$c_{u}\leftarrow argmin_{i}(\mathbf{e}_u,\{c_{1},c_{2},... c_{k}\})$	\\	
	\ForAll{$\{\mathbf{c}\}$}{
		$distance\leftarrow||\mathbf{c}_i-\mathbf{e}_u||_2$ \\
	}

	$\vartriangleright$ \textbf{Cluster Assignment For All Subsequent Epochs} \\
	$\mathbf{e}_u \leftarrow \mathbf{Model}[(\mathbf{m}_u)]$ \tcp*{Compute asynchoronously as the model parameters are received}
	$distance\leftarrow []$\\
	\ForAll{$\{\mathbf{c}\}$}{
		$distance\leftarrow||\mathbf{c}_i-\mathbf{e}_u||_2$ \\
	}
	 $\vartriangleright$ \textbf{Asynchronously Recompute $\{c_{1},c_{2},... c_{k}\}$ Every 2 Global Epochs}\tcp*{using model embeddings available at the server}

\end{algorithm}

Finally, the asynchronous capability of the server allows us to update the semantic sampler model and recompute the effective number of clusters. This is achieved completely independently without incurring any overhead on communication bandwidth between the server and the clients. The dynamic nature of ReFRS architecture enables any clustering algorithm to be substituted.  

\subsection{Aggregator}
Each client model at the server side is assigned a specific cluster by the clustering unit.
The aggregator is responsible for aggregating all the model parameters on each individual cluster and sending the respective aggregated model to the client upon request. When the server receives a client\textquoteright s model parameters, it processes them as explained in Section \ref{Asynchronous_Clustering}. Once the cluster id is obtained, the model parameter and cluster id is fed as input to the aggregator.

\begin{algorithm}
	\caption{Aggregator}
	\label{algo_3}
	\SetKwData{Left}{left}\SetKwData{This}{this}\SetKwData{Up}{up}
	\SetKwFor{ForAll}{for $\forall$}{do in Parallel}{end}
	\SetKwFunction{Union}{Union}\SetKwFunction{FindCompress}{FindCompress}
	\SetKwInOut{Input}{input}\SetKwInOut{Output}{output}
	\KwIn{Client $u$} 
	\KwOut{Aggregated global model $\mathbf{m}_{agg}$}
	\BlankLine
	$c_u\leftarrow$ compute cluster ID for $u$ with Algorithm \ref{algo_2} \\
	$\mathcal{V}_{c_u}\leftarrow$ all existing clients in the $c_u$-th cluster\\
	$\mathbf{m}_{agg}\leftarrow \sum_{u'\in \mathcal{V}_{c_u}}\mathbf{m}_{u'}$\\
	$\mathbf{m}_{agg}\leftarrow \mathbf{m}_{agg}+\mathbf{m}_u$\\
	$\mathbf{m}_{agg}\leftarrow \frac{\mathbf{m}_{agg}}{|\mathcal{V}_{c_u}|+1}$\\
	$\mathcal{V}_{c_u}\leftarrow \mathcal{V}_{c_u} \cup c_u$ \tcp{append the model to its cluster}
	\Return $\mathbf{m}_{agg}$\tcp{aggregated model}
\end{algorithm}

Model parameters, received by the server are aggregated via lines 3-5 of Algorithm \ref{algo_3} and distributed back to the respective clients. In order to asynchronously update clusters on the server side, we keep a copy of the last shared client models. The server updates the clusters after every two global epochs.

\section{Privacy Protection}\label{privacy}
FL assures users some level of privacy since the data is not directly sent to the central server \cite{mcmahan2017communication}. However, it is still possible to infer some information from the shared model parameters by the client. Therefore, a privacy mechanism is still needed to protect user information. In our work, we discuss and demonstrate the compatibility of ReFRS with two widely adopted privacy mechanisms i.e.,  Differentially Private (DP) Stochastic Gradient Descent SGD and Homomorphic Encryption (HE).

\subsection{Differentially Private (DP) SGD}
As a result of its strong privacy guarantees, DP has been widely adopted in FL settings as a privacy mechanism. If some information from a dataset is publicly available, DP can address the privacy leakage of data belonging to an individual. To provide privacy, most DP mechanisms add an independent random noise component to available data \cite{kim2021federated}. By working only on the final parameters, it is possible to protect the privacy of training data. Unfortunately, it may not be possible to directly determine how these parameters are dependent on the training data; if the parameters are overly noisy, the worst-case analysis would undermine the usefulness of the learned model.
We take the sophisticated approach of attempting to control the influence of the training data during the training process, specifically in the SGD computation. Several previous works have adopted this approach (e.g., \cite{song2013stochastic,bassily2014private}). To compute the SGD-DP, each step of the procedure computes the gradient $\nabla_{\theta}\mathcal{L}(\theta, z_i)$ for a random subset of examples, clips the $\ell_2$ norm of each gradient, computes the average, adds noise for privacy, and takes a step in the opposite direction of the average noisy gradient.

\subsection{Homomorphic Encryption (HE)}
As opposed to traditional cryptographic schemes, HE allows certain operations like addition, multiplication, and so forth to be performed directly over encrypted data without the need for decryption. When decrypted, computations produce the same results as if performed on unencrypted data. 
A homomorphic scheme is one that satisfies the following equation:
\begin{equation}
    \Psi(\mathbf{w}_1)*\Psi(\mathbf{w}_2) = \Psi(\mathbf{w}1*\mathbf{w}_2)
\end{equation}
here, $\mathbf{w}_1$ and $\mathbf{w}_2$ are the model parameters from client 1 and 2 at the server side, $*$ represents the mathematical operation performed between the parameters and $\Psi$ represents encryption algorithm. The most commonly $\Psi$ are partially homomorphic encryption (PHE) \cite{morris2013analysis}, somewhat homomorphic (SHE) \cite{boneh2013private}, and fully homomorphic (FHE) \cite{brakerski2014leveled}.  As we need to aggregate encrypted models with a federated setting, we use a fully homomorphic HE scheme (FHE) that offers support for an unbounded number of additive and multiplicative operations. To construct FHE structures, we apply Cheon-Kim-Kim-Song's (CKKS) \cite{cheon2017homomorphic} algorithm, which depends on the hardness of Learning-With-Error (LWE) \cite{regev2009lattices}. With CKKS, we can approximate arithmetic on real and floating point numbers.

\section{Evaluation}\label{eval}
In this section, we evaluate the proposed ReFRS framework in terms of recommendation quality and scalability. Particularly, we aim to investigate the following research questions (RQs) :
 \begin{itemize}
	\item[\textbf{RQ1:}] How effective ReFRS is compared with state-of-the-art federated recommenders?
	\item[\textbf{RQ2:}] Can ReFRS protect user privacy while offering accurate recommendations?
	\item[\textbf{RQ3:}] How is the recommendation quality of ReFRS compared with fully centralized counterparts?
	\item[\textbf{RQ4:}] What are the contributions of the key components of ReFRS?
	\item[\textbf{RQ5:}] How well the asynchronous clustering adjusts to scalability, as more clients are introduced? 
	\item[\textbf{RQ6:}] What are the memory and computation requirements of ReFRS?
\end{itemize}

\subsection{Datasets}
We leverage large-scale datasets where each user would have multiple interactions with items for experimentation. Hence we adopted two publicly available datasets "MovieLens 20M" (ML 20M) \cite{harper2015movielens} and "Last.fm 360K" (FM 2306K) \cite{Celma:Springer2010}. MovieLens 20M contains 20 million interaction records for 162,000 users with a rating density of 0.52\%. Here rating density means "how many items in the dataset were rated by each user". Last.fm dataset contains 17,559,530 interactions (tracks played) by 359,347 users, with a track played density of 0.016\%. Smaller subsets of these datasets have been widely adopted by centralized recommendation literature/research \cite{kang2018self,craw2015music}. The key characteristics of both datasets are summarized in \autoref{tab:dataset}. Using a smaller sample of the original dataset, we evaluate our client's embedding model against state-of-the-art embedding models in a centralized setting. Detailed information about the smaller subsets can be found in \autoref{small_data}, where MovieLens dataset is limited to 100,000 interactions, and Last.fm to 52,551. We first split the datasets into temporally sequential chunks in order to simulate sequential recommendations. Because the models must be deployed on resource-constrained end-devices, we build small sessions, each consisting of 50 interactions.

\begin{table}[h]
	
	\caption{Statistics of the MovieLens 20M (ML) and Last.fm 360K (FM) datasets.}
	\label{tab:dataset}
	\begin{tabular}{llllll} 
		\hline
		Dataset        & Interactions & Users   & Items  & Density & Domain \\ \hline
		ML 20M  & 20,000,263   & 138,493 & 27,278 & 0.52\% & film  \\
		FM 360K    & 17,559,530   & 359,347 & 17,632 & 0.016\% & music\\ \hline
	\end{tabular}
\end{table}
\subsection{Evaluation Criteria}
To efficiently evaluate the performance in our experimentation, we use the widely adopted "leave-one-out" \cite{deshpande2004item} strategy. The latest interaction sequences of each user are held-out for testing, while all previous interaction sequences are used for training. The performance of ranked items is evaluated using Hit Ratio @10 (HR@10) and Normalized Discounted Cumulative Gain @10 (NDCG@10) \cite{wang2010relevance,yu2019generating,sun2020go,chen2020sequence}. In sequential RS settings, HR@10 (\autoref{hit_eq}) is simply the number of times that the ground truth item appears among the top-10 recommendations.
\begin{equation}
	\label{hit_eq}
	HR@10 = \frac{|I^{hit}_{10}|}{|I_{10}|}
\end{equation}

where $|I^{hit}_{10}|$ is the correct number of hits among ten predictions. 
NDCG (\autoref{NDCG}) is simply the normalization of DCG@10 (\autoref{DCG}) score with IDCG@10 (\autoref{IDCG}) score. DCG penalizes relevant items being placed at the bottom and IDCG is the ideal ranking of items in descending order, according their relevance up to position 10. 
\begin{equation}
	\label{DCG}
	DCG@10 = \sum_{i=1}^{I_{10}}\frac{G(i)}{\log_2(i+1)} 
\end{equation}
\begin{equation}
	\label{IDCG}
	IDCG@10 = \sum_{i=1}^{I^{\textbf{*}_{10}}}\frac{G(i)}{\log_2(i+1)} 
\end{equation}
\begin{equation}
	\label{NDCG}
	NDCG@10 = \frac{DCG@10}{IDCG@10}
\end{equation}

where $G$ is the relevance score for item $i$ and  $I^{\textbf{*}_{10}}$ represents the ideal list of items.  We calculate both metrics for each client\textquoteright s test data and report the average score among them.

\subsection{Implementation Details}
Our implementation generally consists of the clients and a central server. The structure of client models is similar for each user and the embedding dimension $d$ for each client is set to 16. The VAE of the client's embedding model consists of two convolution layers \cite{tang2018personalized} with filter sizes of 32, 16 and the kernel size of 2. We restrict the approximation of the posterior using independent gaussians distribution. We test our model by setting the sequence window $w$ of multiple sizes, however since the number of interactions available for most users were low, we keep $w=16$ and the session size is kept at 100. For the recommender at the client side, any state-of-the-art model can be adopted. In our work we used GRU4Rec \cite{hidasi2016parallel} and SASRec \cite{kang2018self} for two reasons. Firstly, they are the most widely adopted SRSs, Secondly, the technologies suggested by these frameworks i.e., GRU and the transformer, are widely adopted in variations by other state-of-the-art SRSs. 

The VAE model used by the semantic sampler, at the server side, consists of two convolution layers (64,32) and three dense layers (16 units each). On the server side standard Gaussian distribution is used.  Using the Elbow method, we determine the optimal number of clusters by setting $k$ between 1 and 100. Initially, 1,000 users were used to train the semantic sampler.

\subsection{Baselines}
We evaluate the performance of ReFRS against three main baselines, using the large-scale datasets. Hyper-parameter settings for these federated baselines frameworks, including ReFRS are given in \autoref{Hyperparameters}.

\begin{table}[t]
	\caption{Hyper-parameters for the baselines and ReFRS. $\beta$ represents the number of epochs before sending model to the server and $\gamma$ represents the learning rate of the clients.}
	\label{Hyperparameters}
	\begin{tabular}{lccccc}
		\hline
		Model   & \multicolumn{1}{l}{\#Clusters} & \multicolumn{1}{l}{Client Epoch} & $\beta$  & $\gamma$    & \multicolumn{1}{l}{Optimization} \\ \hline
		FedAvg  & 1                              & 50                               & 10 & 0.01 & SGD                              \\
		FedFast & 10                             & 50                               & 10 & 0.01 & SGD                              \\
		FRS-1   & 0                              & 50                               & 0  & 0.01 & SGD                              \\
		FRS-2   & 1                              & 50                               & 10 & 0.01 & SGD                              \\
		ReFRS-GRU4Rec   & dynamic                              & 50                               & 10 & 0.01 & SGD                              \\ 
		ReFRS-SASRec   & dynamic                              & 50                               & 10 & 0.01 & SGD                              \\
		\hline
	\end{tabular}
\end{table}

\begin{itemize}
	\item \textbf{FedAvg} \cite{sattler2019robust}: The popular FedAvg algorithm aggregates the parameters of all client models into a single global model. In this work, each client model consists of a General Matrix Factorization (GMF) model \cite{he2017neural}, as adopted by FedFast \cite{muhammad2020fedfast}.
	
	\item \textbf{FedFast} \cite{muhammad2020fedfast}: This method accelerates FRS tasks by applying an active aggregation method that propagates the updated models to similar clients. Each client contains multiple users, which are clustered based upon user similarity information. GMF \cite{he2017neural} is used for the client model.
	
	\item \textbf{Federated Recommendation System-1 (FRS-1)} : FRS-1 is the basic setting of ReFRS. It consists of a client model, using GRU4Rec as a recommender. Here, the client learns recommendations for each user separately, without the guidance of a federated server.
	
	\item \textbf{Federated Recommendation System-2 (FRS-2)} : FRS-2 is the second basic setting of ReFRS, in which the federated server adopts FedAvg to aggregate shared models into a single global model. The client uses GRU4Rec as a recommender in FRS-2.
	
	\item \textbf{FedAvg-GRU4Rec} : This setting adopts FedAvg algorithm to enable decentralized learning of GRU4Rec \cite{hidasi2015session} algorithm.  GRU4Rec utilizes GRU layers to model user interaction sequences per session.
	
	\item \textbf{FedAvg-SASRec} : This setting adopts FedAvg algorithm to enable decentralized learning of SASRec \cite{kang2018self} algorithm.  SASRec uses multi-head attention in order to recommend the next item.
	
	\item \textbf{FedFast-GRU4Rec} : This setting of FedFast adopts GRU4Rec as the recommender model.
	
	\item \textbf{FedFast-SASRec} : This setting of FedFast adopts SASRec as the recommender model.
	
	\item \textbf{ReFRS-GRU4Rec} : 
	This setting of ReFRS adopts GRU4Rec to compute next item recommendation. 
	
	\item \textbf{ReFRS-SASRec} : This setting of ReFRS adopts SASRec as the recommender model. 
	
\end{itemize}


To show the effectiveness of ReFRS's client module, we also compare it against the following state-of-the-art centralized sequential recommender models, using smaller datasets. All hyper-parameters are set following the original papers.	
\begin{itemize}
	\item \textbf{AutoInt} \cite{song2019autoint} learns feature interactions using the multi-head self-attentive neural network
	\item \textbf{GRU4Rec} \cite{hidasi2015session} utilizes GRU layers to model user interaction sequences per session.
	\item \textbf{Caser} \cite{tang2018personalized} uses horizontal and vertical convolutional to capture high-order Markov Chains.
	\item \textbf{SASRec} \cite{kang2018self} uses multi-head attention in order recommend next item, for sequential recommendation task.
	\item \textbf{BERT4Rec} \cite{sun2019bert4rec} adopts a Cloze objective loss, on top of bidirectional self-attention mechanism, for sequential
	recommendation task.
	\item \textbf{HGN} \cite{ma2019hierarchical}  uses hierarchical gated networks for capturing both long-term and short-term user interests.
	\item \textbf{FDSA} \cite{zhang2019feature} generates feature sequences and model\textquoteright s feature transition patterns using a feature level self-attention block.
	\item $\textbf{S}^3$\textbf{-Rec} \cite{zhou2020s3} utilize the intrinsic data correlation to derive self-supervision signals and enhance the data representations via per-training. All hyper-parameters are set following the suggestions from the original papers.	
	
\end{itemize}

\subsection{RQ-1: Performance Comparison on Large-scale Datasets}

\begin{table}[]
	\caption{HR@10 and NDCG@10 of FedAvg, FedFast, FedGRU4Rec, FedSASRec, ReFRS-GRU4Rec and ReFRS-SASRec for different subsets of MovieLens 20M and Last.fm 360K datasets.}
	\label{performance_table}
	\scalebox{0.82}{
	\begin{tabular}{ll|cc|cc|cc|cc}
		\hline
		&         & \textbf{FedAvg} & \textbf{FedFast} & \multicolumn{2}{c|}{\textbf{FedAvg}} & \multicolumn{2}{c|}{\textbf{FedFast}} & \multicolumn{2}{c}{\textbf{ReFRS}} \\
		\textbf{Dataset}  &         & \textbf{GMF}    & \textbf{GMF}     & \textbf{GRU4Rec}      & \textbf{SASRec}      & \textbf{GRU4Rec}       & \textbf{SASRec}      & \textbf{GRU4Rec}      & \textbf{SASRec}     \\ \hline
		ML-10\%  & HR@10   & 0.4526 & 0.5612  & 0.4930  & 0.5298 & 0.5733  & 0.6160 & 0.6164  & 0.6624 \\
		& NDCG@10 & 0.2716 & 0.3404  & 0.3017  & 0.4047 & 0.3508  & 0.4706 & 0.3772  & 0.5060 \\\hline
		ML-20\%  & HR@10   & 0.3998 & 0.5520  & 0.4488  & 0.4856 & 0.5219  & 0.5647 & 0.5612  & 0.6072 \\
		& NDCG@10 & 0.1811 & 0.2576  & 0.2796  & 0.3605 & 0.3251  & 0.4192 & 0.3496  & 0.4508 \\\hline
		ML-30\%  & HR@10   & 0.3168 & 0.4784  & 0.3973  & 0.4341 & 0.4620  & 0.5048 & 0.4968  & 0.5428 \\
		& NDCG@10 & 0.1433 & 0.2116  & 0.2281  & 0.2943 & 0.2652  & 0.3422 & 0.2852  & 0.3680 \\\hline
		ML-40\%  & HR@10   & 0.1735 & 0.3220  & 0.3017  & 0.3458 & 0.3508  & 0.4021 & 0.3772  & 0.4324 \\
		& NDCG@10 & 0.0830 & 0.1288  & 0.1913  & 0.2649 & 0.2225  & 0.3080 & 0.2392  & 0.3312 \\\hline
		ML-50\%  & HR@10   & 0.0905 & 0.1932  & 0.2207  & 0.2649 & 0.2567  & 0.3080 & 0.2760  & 0.3312 \\
		& NDCG    & 0.0528 & 0.0920  & 0.1104  & 0.1619 & 0.1283  & 0.1882 & 0.1380  & 0.2024 \\\hline
		ML-100\% & HR@10   & 0.0604 & 0.1564  & 0.1692  & 0.1987 & 0.1968  & 0.2310 & 0.2116  & 0.2484 \\
		& NDCG@10 & 0.0075 & 0.0460  & 0.0662  & 0.0957 & 0.0770  & 0.1112 & 0.0828  & 0.1196 \\\hline
		FM-10\%  & HR@10   & 0.4602 & 0.5704  & 0.4562  & 0.5077 & 0.5305  & 0.5904 & 0.5704  & 0.6348 \\
		& NDCG@10 & 0.3093 & 0.4140  & 0.3311  & 0.4121 & 0.3850  & 0.4791 & 0.4140  & 0.5152 \\\hline
		FM-20\%  & HR@10   & 0.2490 & 0.3588  & 0.3900  & 0.4341 & 0.4535  & 0.5048 & 0.4876  & 0.5428 \\
		& NDCG@10 & 0.1358 & 0.2116  & 0.2649  & 0.3017 & 0.3080  & 0.3508 & 0.3312  & 0.3772 \\\hline
		FM-30\%  & HR@10   & 0.1961 & 0.2944  & 0.3753  & 0.4047 & 0.4364  & 0.4706 & 0.4692  & 0.5060 \\
		& NDCG@10 & 0.1132 & 0.1564  & 0.2428  & 0.2870 & 0.2823  & 0.3337 & 0.3036  & 0.3588 \\\hline
		FM-40\%  & HR@10   & 0.1811 & 0.2576  & 0.2796  & 0.3017 & 0.3251  & 0.3508 & 0.3496  & 0.3772 \\
		& NDCG@10 & 0.0453 & 0.1012  & 0.1545  & 0.1692 & 0.1797  & 0.1968 & 0.1932  & 0.2116 \\\hline
		FM-50\%  & HR@10   & 0.1282 & 0.1932  & 0.2281  & 0.2428 & 0.2652  & 0.2823 & 0.2852  & 0.3036 \\
		& NDCG@10 & 0.0075 & 0.0552  & 0.1104  & 0.1472 & 0.1283  & 0.1711 & 0.1380  & 0.1840 \\\hline
		FM-100\% & HR@10   & 0.0604 & 0.1380  & 0.1251  & 0.1324 & 0.1455  & 0.1540 & 0.1564  & 0.1656 \\
		& NDCG@10 & 0.0000 & 0.0092  & 0.0589  & 0.0809 & 0.0684  & 0.0941 & 0.0736  & 0.1012   \\ \hline
	\end{tabular}}
\end{table}

We evaluate the performance of ReFRS in federated settings against FedAvg and FedFast.
In order the effectiveness of our embedding method we upscale the basic General Matrix Factorization (GMF) based recommender approch adopted by FedFast and FedAvg. We replace basic GMF model by state-of-the-art GRU4Rec \cite{hidasi2015session} and SASRec \cite{kang2018self} models. In our work we use GRU4Rec and SASRec since they are most widely adopted SRSs. In addition the learning mechanism proposed by these frameworks i.e., GRU and the transformer, are widly adopted in different variations by other state-of-the-art SRS.

In contrast to the FedFast paper, \cite{muhammad2020fedfast}, which uses a small-scale dataset to evaluate their semantic models, in our work we use datasets that are over 200 percent larger. This approach tests the scalability of the proposed frameworks and simulates the real world better.
This experiment evaluates the performance of all baselines as the number of client devices are increased. We divide ML and FM datasets into subsets containing 10\%, 20\%, 30\%, 40\%, 50\% and 100\% of the actual users. These subsets are generated at random. The HR@10 and NDCG@10 scores of ReFRS against all the base lines are given in \autoref{performance_table}.

In \autoref{performance_table} all settings of the  FedAvg algorithm constantly under performs our proposed approach. This is because, a large number of users in our datasets do not have enough interactions for the RS to effectively predict user's long term interests and change in interaction patterns. This lack of interaction data in a large number of users causes lower performance of the global model as all user models are aggregated together. Global generalization of models improves the performance in the case of FedFast. However, since their model generalization is based upon aggregating models of users with similar Bios, they fail to outperform both settings ReFRS which groups users based upon model parameters. This shows that the semantic sampler in ReFRS maps similarly behaving user models close in the embedding space. ReFRS with SASRec performs better than GRU4Rec, indicating that attention-based techniques more accurately synthesize user behavior from embeddings, than recurrent neural networks. 
Note, shared model gradients do not provide any information regarding user behavior, rather each gradients only carry some effect on the input sequence \cite{sundararajan2017axiomatic} (models with similar gradients will have the same effect). Hence, no personal information is leaked at the server side.



\subsection{RQ-2: Effects of Privacy on Performance}

FL assures users some level of privacy since the data is not directly sent to the central server \cite{mcmahan2017communication}. However, it is still possible to infer some information from the shared model parameters by the client. This
may lead to the leakage of sensitive or private information. For this reason, to the effect of different privacy mechanisms on ReFRS, we adopt two widely used privacy mechanisms i.e.,  Differentially Private (DP) Stochastic Gradient Descent SGD and Homomorphic Encryption (HE). 
The DP noise multiplier is adjusted to regulate the amount of noise added during training. As a general rule, more noise corresponds to better privacy and less utility. In our experiments and to obtain rigorous privacy guarantees \cite{balle2018improving}, the value is set to 0.3. We evaluate the performance of ReFRS-GRU4Rec and ReFRS-SASRec using both encryption mechanisms.

\begin{table}[]
	\caption{HR@10 and NDCG@10 of ReFRS-GRU4Rec and ReFRS-SASRec, using DP-SGD and HP, for different subsets of MovieLens 20M and Last.fm 360K datasets.}
	\label{encryp}
	\scalebox{0.82}{
	\begin{tabular}{ll|cc|cc}
		\hline
		& & \multicolumn{2}{c|}{\textbf{DP-SGD}}           & \multicolumn{2}{c}{\textbf{HE}}                \\ \cline{3-6}
		\multicolumn{1}{c}{\textbf{Dataset}} &    \multicolumn{1}{c|}{\textbf{Matrices}}                                    & \textbf{ReFRS-GRU4Rec} & \textbf{ReFRS-SASRec} & \textbf{ReFRS-GRU4Rec} & \textbf{ReFRS-SASRec} \\  \hline 
		ML-10\%                              & HR@10                                  & 0.4095                 & 0.4459                & 0.6164                 & 0.6624                \\
		& NDCG@10                                & 0.2366                 & 0.2821                & 0.3772                 & 0.5060                \\ \hline
		ML-20\%                              & HR@10                                  & 0.3640                 & 0.4186                & 0.5612                 & 0.6072                \\
		& NDCG@10                                & 0.2912                 & 0.3367                & 0.3496                 & 0.4508                \\ \hline
		ML-30\%                              & HR@10                                  & 0.3185                 & 0.3549                & 0.4968                 & 0.5428                \\
		& NDCG@10                                & 0.2366                 & 0.2639                & 0.2852                 & 0.3680                \\ \hline
		ML-40\%                              & HR@10                                  & 0.3003                 & 0.3458                & 0.3772                 & 0.4324                \\
		& NDCG@10                                & 0.1729                 & 0.1911                & 0.2392                 & 0.3312                \\ \hline
		ML-50\%                              & HR@10                                  & 0.2275                 & 0.2548                & 0.2760                 & 0.3312                \\
		& NDCG                                   & 0.0819                 & 0.1001                & 0.1380                 & 0.2024                \\ \hline
		ML-100\%                             & HR@10                                  & 0.1911                 & 0.1911                & 0.2116                 & 0.2484                \\
		& NDCG@10                                & 0.0455                 & 0.0637                & 0.0828                 & 0.1196                \\ \hline
		FM-10\%                              & HR@10                                  & 0.5005                 & 0.5278                & 0.5704                 & 0.6348                \\
		& NDCG@10                                & 0.3094                 & 0.3367                & 0.4140                 & 0.5152                \\ \hline
		FM-20\%                              & HR@10                                  & 0.4277                 & 0.4641                & 0.4876                 & 0.5428                \\
		& NDCG@10                                & 0.2548                 & 0.2912                & 0.3312                 & 0.3772                \\ \hline
		FM-30\%                              & HR@10                                  & 0.3731                 & 0.3913                & 0.4692                 & 0.5060                \\
		& NDCG@10                                & 0.2093                 & 0.2275                & 0.3036                 & 0.3588                \\ \hline
		FM-40\%                              & HR@10                                  & 0.3003                 & 0.3367                & 0.3496                 & 0.3772                \\
		& NDCG@10                                & 0.1274                 & 0.1638                & 0.1932                 & 0.2116                \\ \hline
		FM-50\%                              & HR@10                                  & 0.2366                 & 0.2639                & 0.2852                 & 0.3036                \\
		& NDCG@10                                & 0.0728                 & 0.1183                & 0.1380                 & 0.1840                \\ \hline
		FM-100\%                             & HR@10                                  & 0.1183                 & 0.1365                & 0.1564                 & 0.1656                \\
		& NDCG@10                                & 0.0546                 & 0.0637                & 0.0736                 & 0.1012                \\ \hline
	\end{tabular}}
\end{table}

\autoref{encryp} shows the effect of DP-SGD and HE on our framework. The results show degradation in the results when DP-SGD is used. This is because the inclusion of noise during training of the client model in DP-SGD, makes the reconstruction task inaccurate. In addition, the noisy model parameter makes the semantic server improperly group client models. Therefore, the model fails to capture valuable client interaction patterns. However, in the case of HE a secure message is sent to the server, on which mathematical operations can be performed precisely. This does not effect either the client model or the semantic server. Hence, we suggest HE should be adopted as encryption mechanism when semantic sampling is evolved at the server side. The key sharing strategy adopted here is the same as in federated learning literature i.e. the private key distributed to each client is the same, distributed by a third party (unknown by the server and top-level clients).

\subsection{RQ-3: Effectiveness against Centralized Models}
	ReFRS is designed for federated settings, where the central server has no user interaction data available. However, in order to further demonstrate the effectiveness of ReFRS's client model, we compare it against state-of-the-art centralized SRSs. For this experiment we test the recommendation performance of ReFRS using SASRec as the recommender model of our client model.
We select two smaller subsets of the previously adopted datasets i.e., MovieLens 100K (ML 100K) and Last.fm 1K (FM 1K) and evaluate performance of ReFRS against state state-of-the-art centralized  SRSs. ML 100K contains 100,000 interactions of 1,682 users on 943 items, while FM 1K have 52,551 interactions of 1,090 users on 3,646 items. 

The  statistics of the datasets are given in Table \ref{small_data}.
\begin{table}[H]
	\caption{Descriptive statistics of the MovieLens100k and Last.fm 1k datasets.}
	\label{small_data}
	\begin{tabular}{llll}
		\hline
		Dataset    & Users  & Items  & Interactions \\ \hline
		MovieLens 100K       & 1,682 & 943 & 100,000      \\
		Last.fm 1k & 1,090  & 3,646  & 52,551       \\ \hline
	\end{tabular}
\end{table}

\begin{table}[]
	\caption{Performance evaluation of ReFRS against state-of-the-art centralized sequential recommender systems on small-scale datasets. \textbf{Bold*} number identifies the best performing model.}
	\label{small_scale_r}
	\scalebox{0.74}{
		\begin{tabular}{l|lccccccHccc}
			\hline
			Datasets                                                                & Matrix  & AutoInt & GRU4Rec & Caser  & SASRec & BERT4Rec & HGN    & GRU4RecF & FDSA   & $S^3$-Rec & ReFRS  \\ \hline
			\multirow{5}{*}{\begin{tabular}[c]{@{}l@{}}ML\\ 100K\end{tabular}} & HR@1    & 0.0731  & 0.2092  & 0.2194 & 0.2351 & {0.2863}   & 0.2428 & 0.2092   & 0.2198 & 0.2091 & \textbf{0.3001} \\
			& HR@5    & 0.2249  & 0.5103  & 0.5353 & 0.5434 & 0.5876   & 0.5768 & 0.5103   & 0.5728 & {0.5885} & \textbf{0.5897} \\
			& HR@10   & 0.3367  & 0.6351  & 0.6692 & 0.6629 & 0.697    & 0.7411 & 0.6351   & 0.7555 & 0.6725 & \textbf{0.7565} \\
			& NDCG@5  & 0.1501  & 0.3705  & 0.3832 & 0.398  & {0.4454}   & 0.4162 & 0.3705   & 0.4014 & 0.3401 & \textbf{0.4522} \\
			& NDCG@10 & 0.186   & 0.4064  & 0.4268 & 0.4368 & {0.4818}   & 0.4695 & 0.4064   & 0.4607 & 0.3934 & \textbf{0.4851} \\ \hline
			\multirow{5}{*}{\begin{tabular}[c]{@{}l@{}}FM\\ 1k\end{tabular}}   & HR@1    & 0.0349  & 0.0642  & 0.0899 & 0.1211 & 0.122    & 0.0908 & {0.1385}   & 0.0936 &{ 0.1743} & \textbf{0.1753}\\
			& HR@5    & 0.155   & 0.1817  & 0.2982 & 0.3385 & {0.3569}   & 0.2872 & 0.3202   & 0.2624 & {0.4523} & \textbf{0.4530}\\
			
			& HR@10   & 0.2596  & 0.2817  & 0.4431 & 0.4706 & 0.4991   & 0.4193 & 0.467    & 0.4055 & \textbf{0.5535} & {0.5447} \\
			
			& NDCG@5  & 0.0946  & 0.1228  & 0.196  & 0.233  & 0.2409   & 0.1896 & 0.2301   & 0.1766 & {0.3156} & \textbf{0.3174} \\
			
			& NDCG@10 & 0.1285  & 0.155   & 0.2428 & 0.2755 & 0.2871   & 0.2324 & 0.2775   & 0.2225 & {0.3583} & \textbf{0.3566} \\ \hline
		\end{tabular}
	}
\end{table}


\autoref{small_scale_r} shows that our model performs comparably to state-of-the-art centralized SRSs. This is because the recommender model on each mobile device is personalized, which can effectively address the issues of biases in the centralized models.  Central models
are often critiqued as being “single thought” models that favors only the users with rich interaction data, leading to unsatisfactory performance for the long-tail users.  Traditional centralized machine learning relies mainly on generic models: models tuned to an average target population. However, the 'good' performance by these generic models does not necessarily translate to each individual.
For MovieLens dataset, BERT4Rec performs comparably to our model. This shows that our model can also capture repeated behaviours (rating similar genre movies) of users in the ML dataset. For the Last.fm dataset, where the users' interest evolve over time, ReFRS along with $S^3$-Rec are able to effectively capture higher-order interaction behaviors.

\subsection{RQ-4: Ablation Study of the key components of ReFRS}

\subsubsection{Ablation Study of Client Module}
This study aims to demonstrate how the ablation of the client model affects the accuracy and size of the model.
We perform the ablation study of our proposed embedding (client) module by removing various layers from the proposed generative model and testing the Hit@10 ratio. For this study, we compare the embedding model of our proposed ReFRS, which has VQ-VAE reinforced by a multi-head attention layer, against a simple VAE model and VQ-VAE model. For consistent experimentation, we test all models on both MovieLens and Last.fm datasets and report the ablated models' performance as their average. 

\begin{figure}[h]
	\centering
	\subfloat[][\centering Hit@10 ratio on MovieLens 20M dataset.]{{\includegraphics[width=6.5cm]{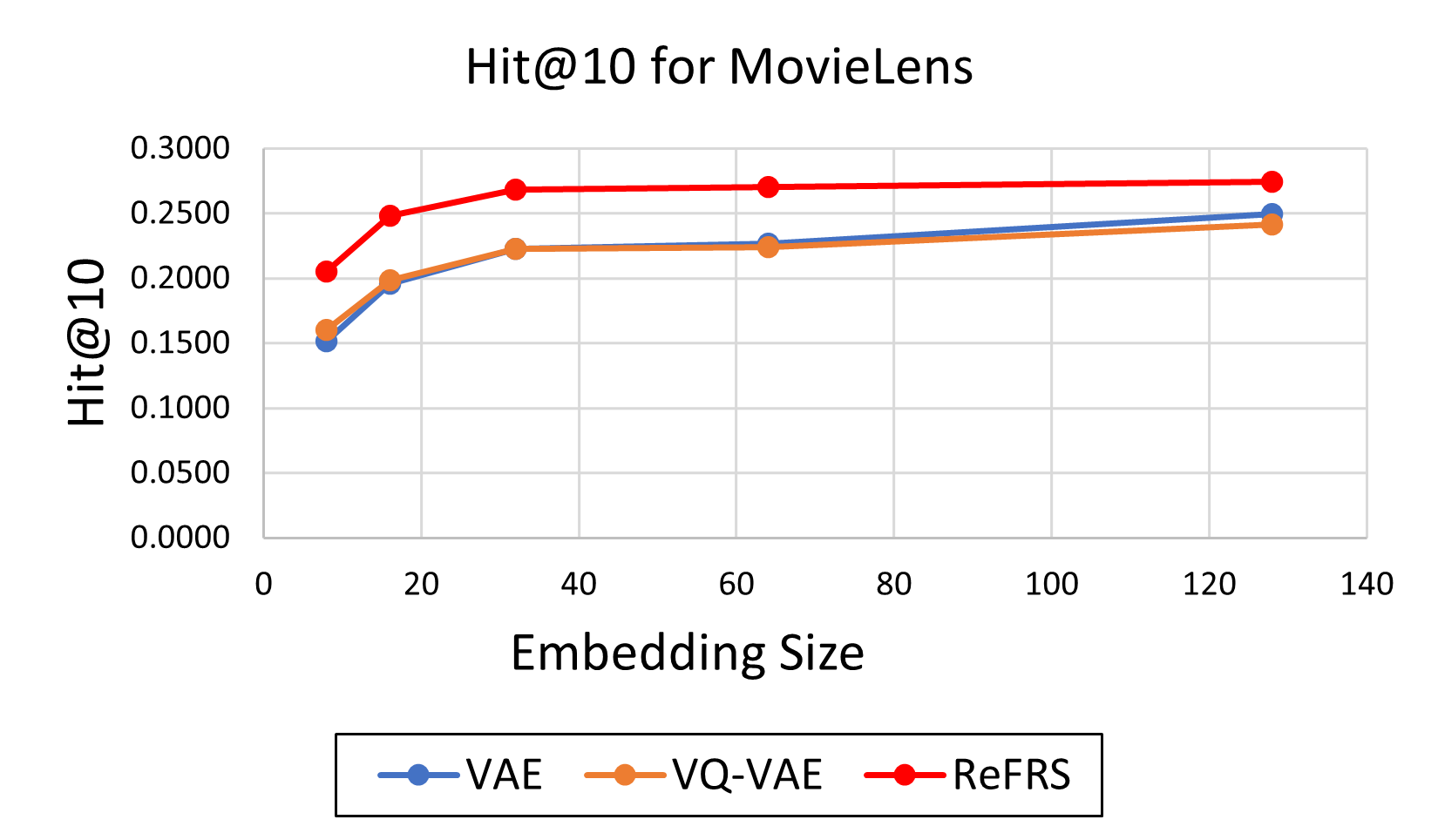} }}
	\qquad
	\subfloat[][\centering Hit@10 ratio on Last.fm 360K dataset.]{{\includegraphics[width=6.5cm]{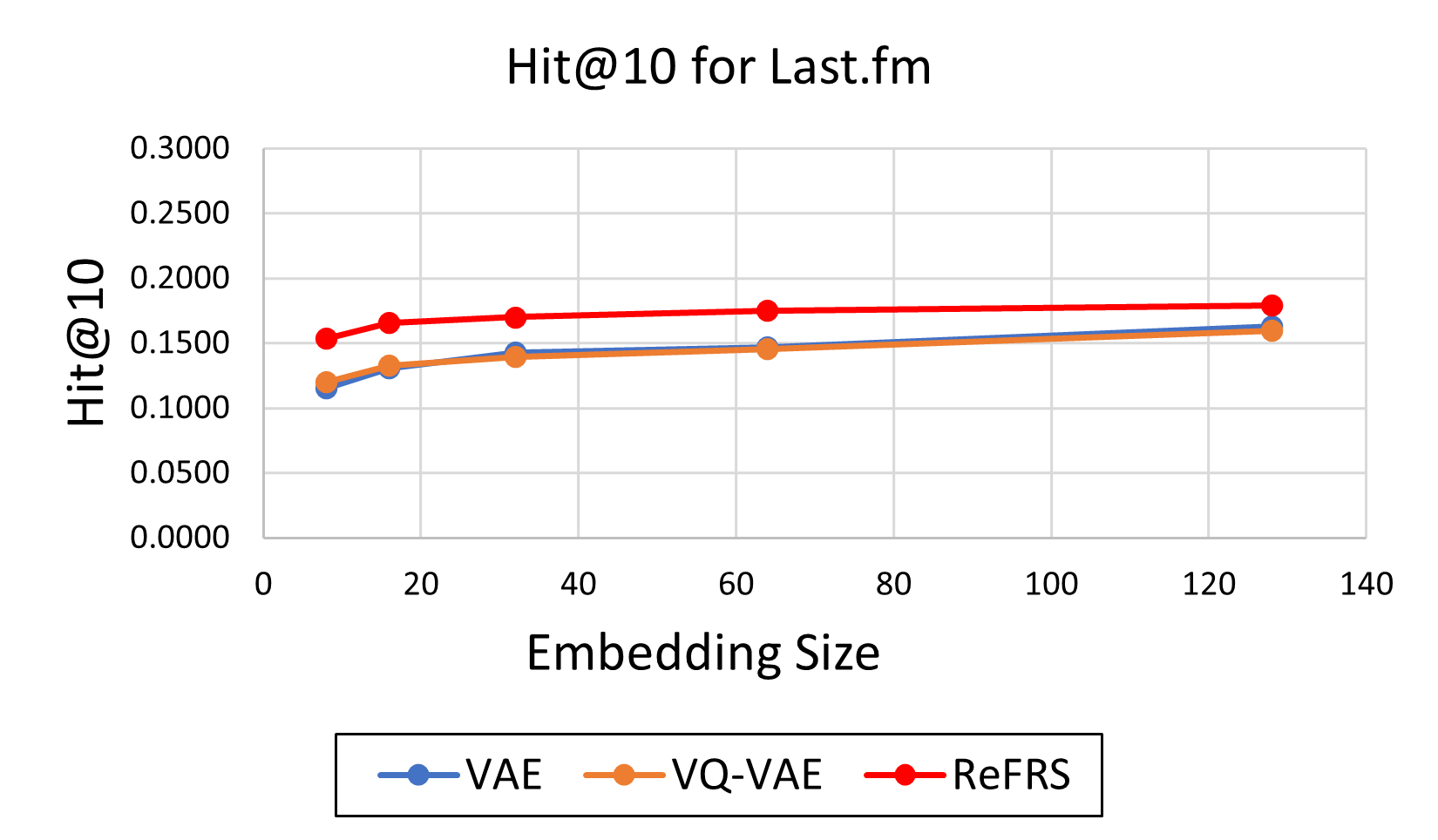} }}
	\qquad
	\subfloat[][\centering Space in KBs taken by each proposed model in memory, by alternating the embedding size.]{{\includegraphics[width=6.5cm]{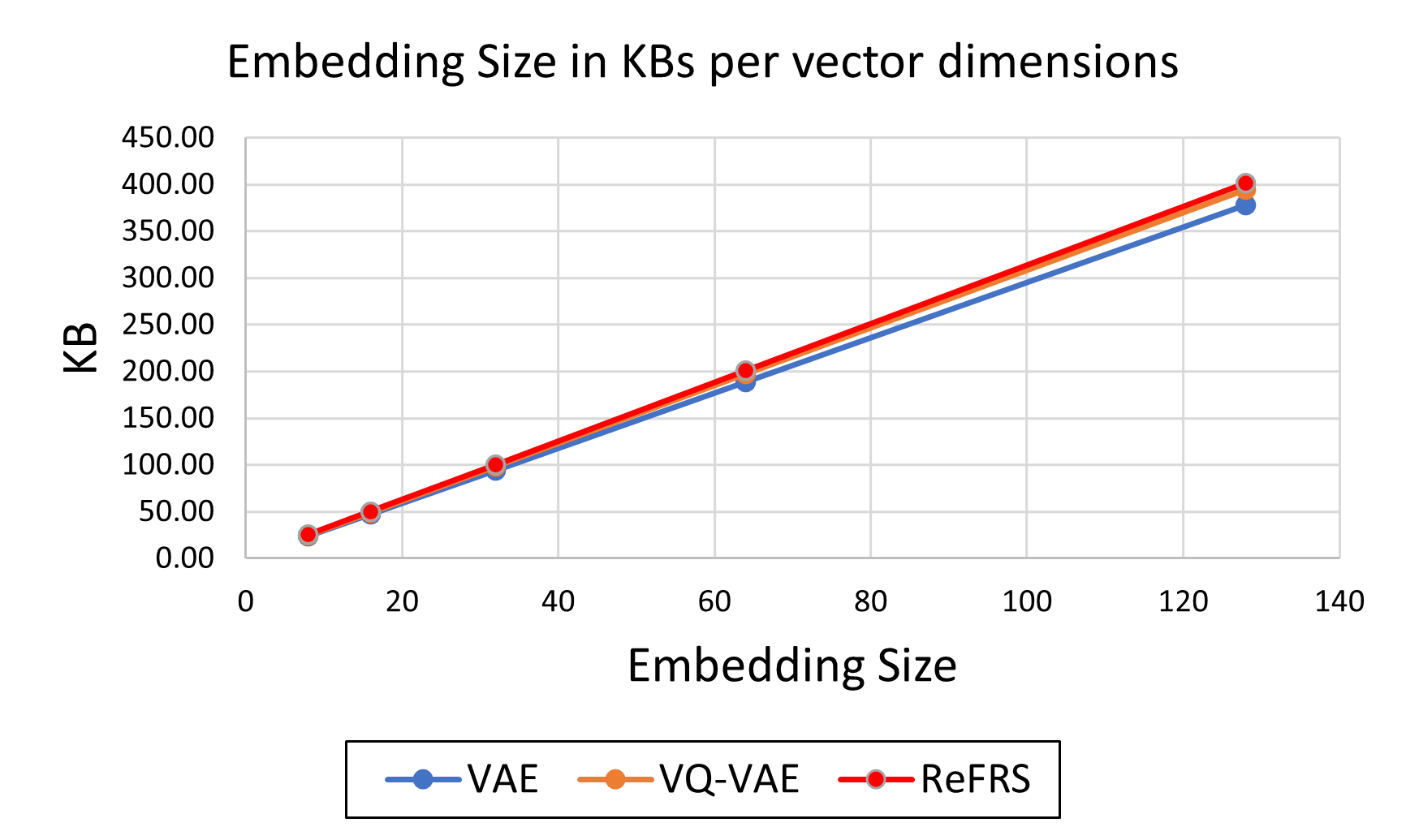} }}
	\caption{(a) and (b) are the ablation study of the embedding model of our proposed ReFRS model, by comparing its embedding model with a simple VAE, VQ-VAE and our proposed VQ-VAE with multi-head attention (ReFRS). We test the performance of each model on different embedding sizes. (c) shows the space (in KBs) taken by each proposed model in memory, by alternating the embedding size.}
	\label{emb_ablate}
\end{figure}

As depicted in \autoref{emb_ablate}(a) and \autoref{emb_ablate}(b), our proposed attention guided embedding method outperforms simple VAE and VQ-VAE. This is because our method produces guided embeddings, which enhance similar and repeated interactions and diminish arbitrary ones.  \autoref{emb_ablate}(a) and \autoref{emb_ablate}(b) also show that the Hit@10 starts to stabilize when the embedding size reaches 16. As a result, our model's embedding size was set to 16 to meet the resource-efficient requirement.
We can see from the \autoref{emb_ablate}(c) that adding VQ and attention layers with (16) embedding size does not lead to a considerable memory overhead, as compared to baselines.

\subsubsection{Ablation Study of Server Module}
Note that unlike GMF-based FedFast or FedAvg, ReFRS operates in a sequential fashion. Hence, we divide the input streams into sessions of new interactions, added every global epoch.
To verify the effectiveness and efficiency of the federated component at each global epoch, we carry out ablation study of ReFRS, on the same large-scale datasets.

\begin{figure}[]
	\centering
	\includegraphics[width=\linewidth]{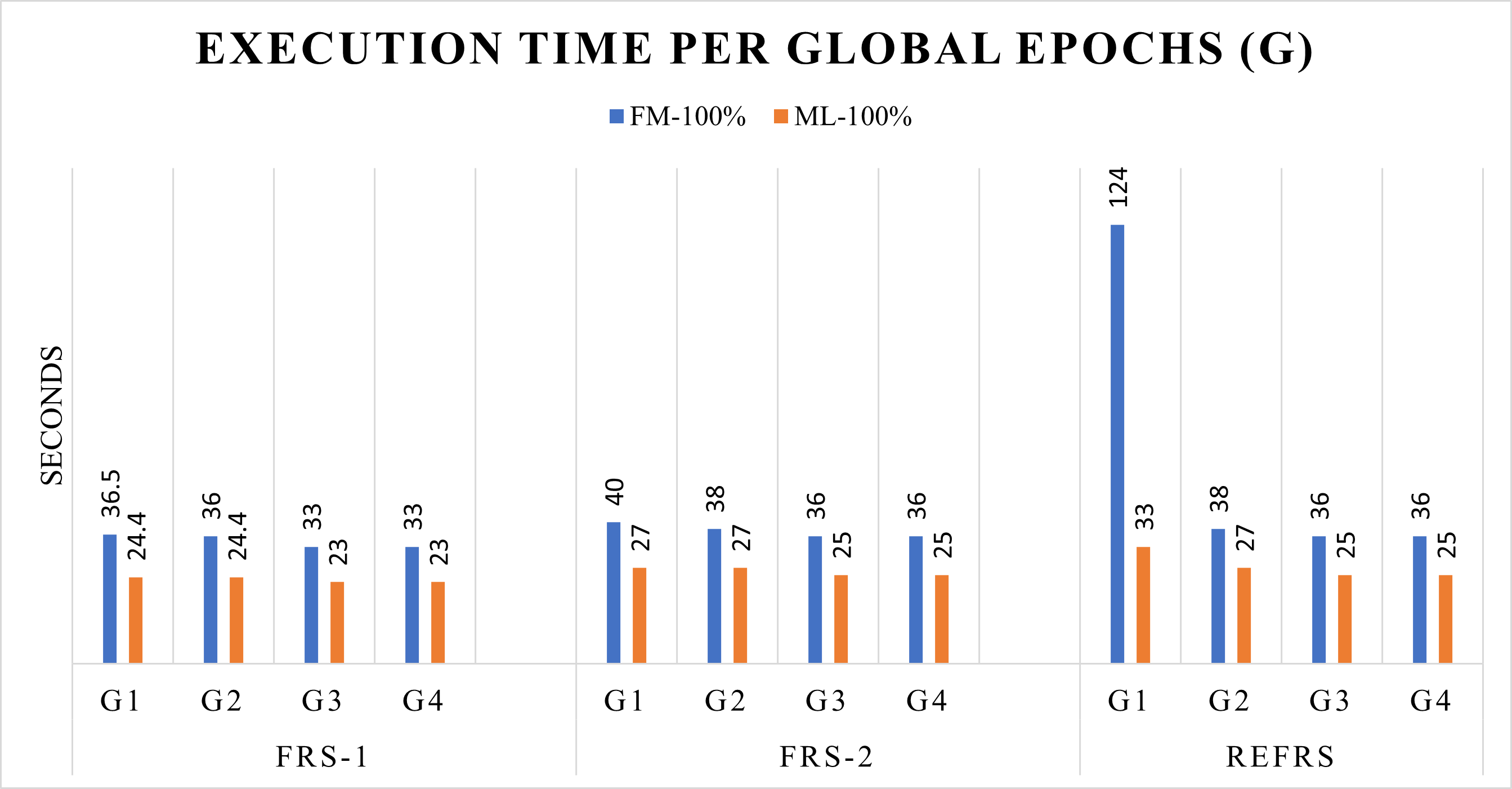}
	\caption{Time efficiency (in seconds) of ReFRS in against FRS-1 and FRS-2, per Global Epoch (G).}
	\label{time_fig}
\end{figure}

We test the effectiveness and efficiency of our proposed semantic similarity based grouping against FRS-1, which is the basic setting of ReFRS with  federated component removed and FRS-2 which aggregates all the client models using simple FedAvg.


To show comparability of neural federated system of ReFRS in efficiency, against FRS-1 and FRS-2, we train each framework for four  global epochs, on four sessions of data. Global updates in \autoref{time_fig} represent the time taken by the slowest worker to converge for five local training session, updating to the federated server and receiving the aggregated model back on the client. For each user, this involves feature sampling, embedding client models, recommending times, and server aggregation.

\begin{figure}[]
	\centering
	\includegraphics[width=\linewidth]{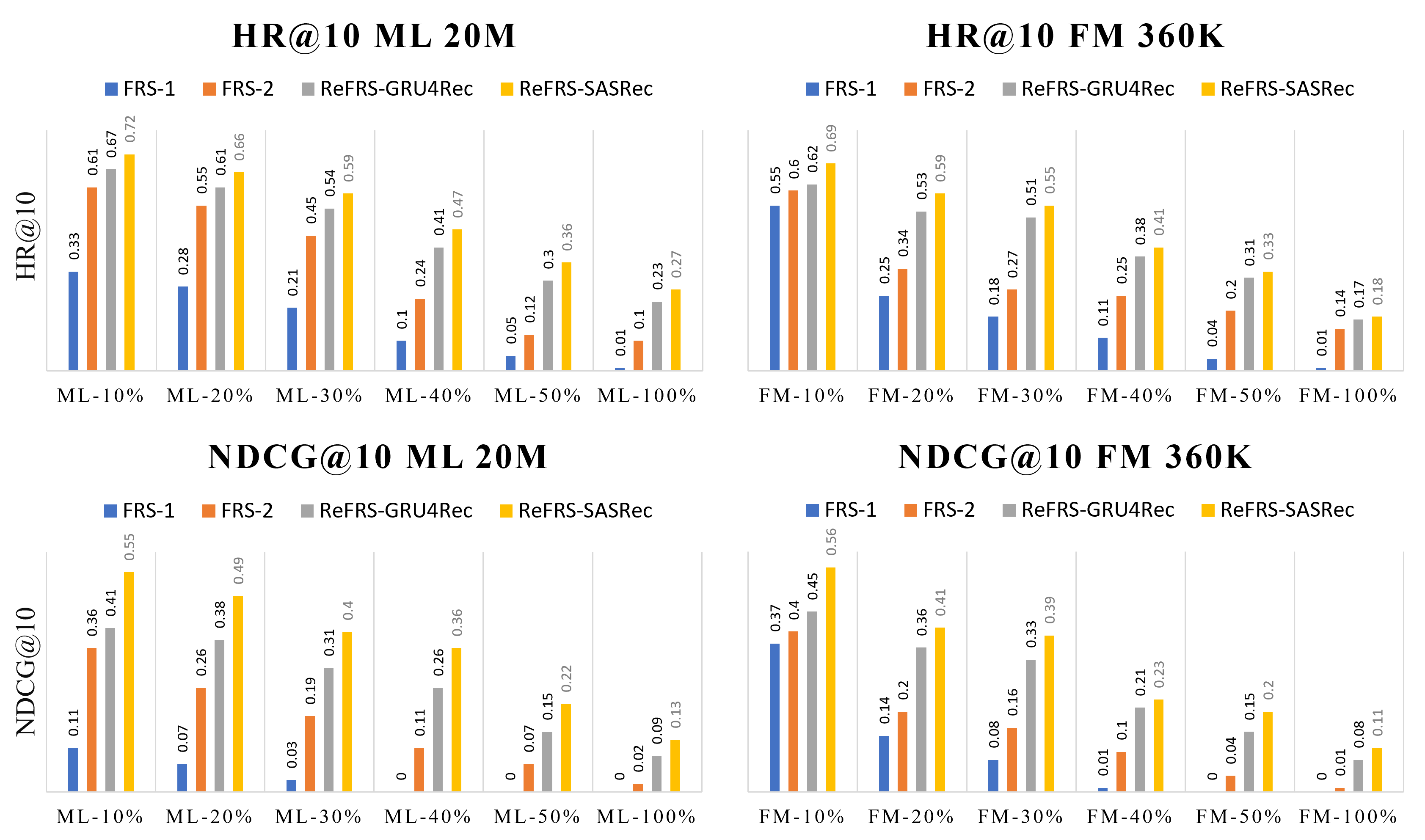}
	\caption{Ablation study of our ReFRS on multiple subsets of ML 20M and FM 360K dataset. Bar peaks represents HR@10.}
	\label{fig:ablation}
\end{figure}

For FRS-1, global update represents convergence of the slowest local model for five local training session. In \autoref{time_fig} we can see that after the first global update, where the global model is trained from scratch, the time efficiency for all there epochs is comparable with FRS-2. This is because the global model is updated asynchronously after its initialization during the first epoch.
We also compare the quality of these settings by altering the number of users in training. As shown in \autoref{performance_table} and  \autoref{fig:ablation}, although FRS-1 has a quicker training time, ReFRS outperforms both FRS-1 and FRS-2 for recommendation quality. These results clearly indicate the  importance of diversity in the recommendation process, which is key objective of ReFRS. FM dataset takes more training time because of the larger amount of input features of items.


\subsection{RQ-5: Server Scalability}
\begin{figure}[t]
	\centering
	\includegraphics[width=\linewidth]{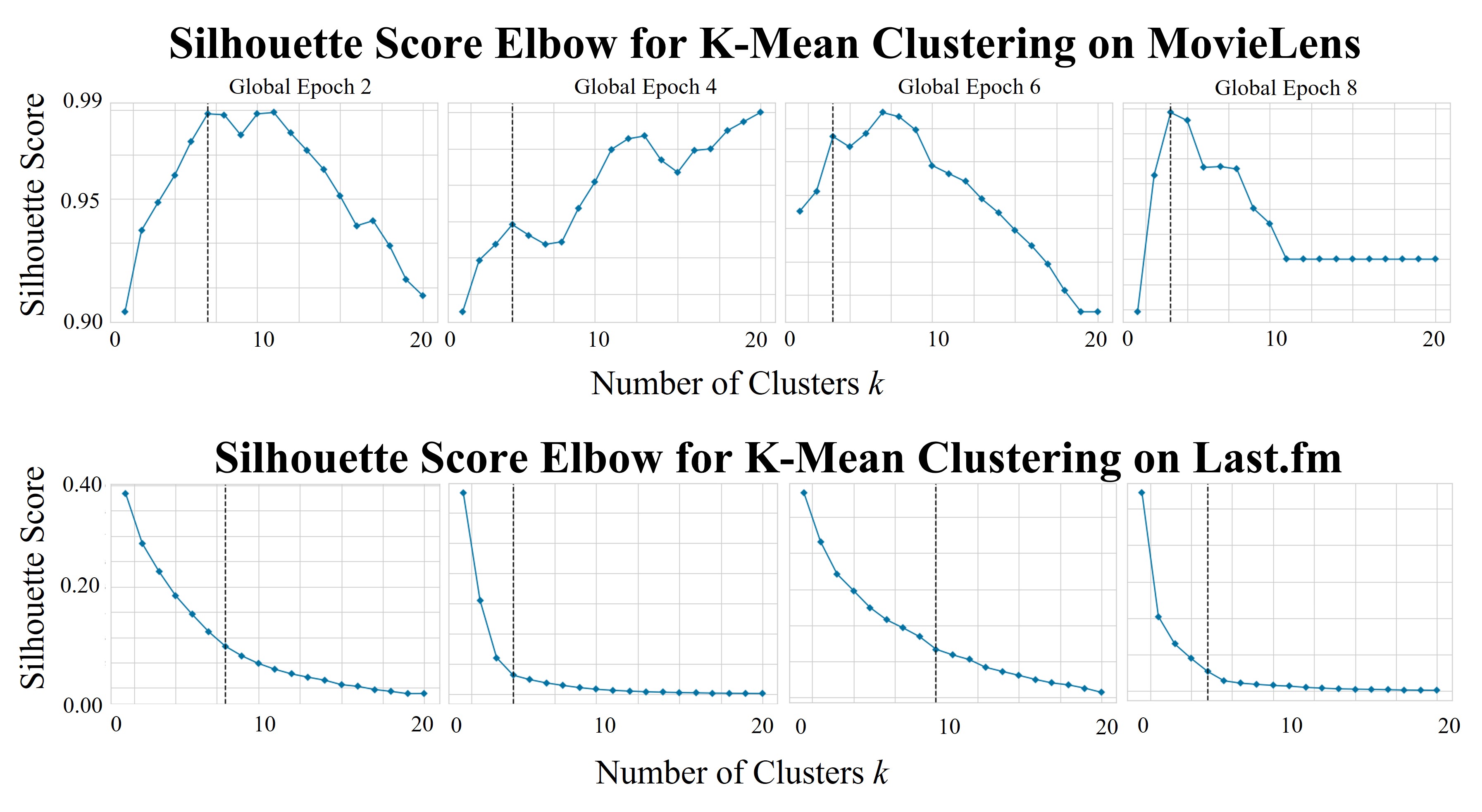}
	\caption{\textit{Top}: Number of clusters formed for MovieLens 20M dataset as the global update enhances.  
		\textit{Bottom}: Number of clusters formed for Last.fm 360K dataset as the global update enhances.}
	\label{fig:cluster}
\end{figure}

In real-time simulation settings, input into the client model are the next stream of user interactions. With the inclusion of new user interactions or new user clients, global groupings of client models are updated. Figure \ref{fig:cluster} shows the dynamic update of user grouping as the latest sequences of interactions for each user are added. Changes in the number of clusters over time by the ReFRS server depicts its adaptability to change and acceptance to scalability.  ReFRS dynamically caters new interactions, as well as introduction of new clients. 

\subsection{RQ-6: Memory and Computation Requirements of ReFRS}

\begin{table}[t]
	\caption{Average memory consumption of FedAvg, FedFast and ReFRS on a single client. The computation time represents the time taken by the client's for taring single local batch.}
	\label{mem}
	\begin{tabular}{l|c|c|c|}
		\cline{2-4}
		& \textbf{FedAvg}        & \textbf{FedFast}      & \textbf{ReFRS}             \\ \hline
		Number of Embedded Vectors Stored        & 18K-28K       & 18K-28K       & 100               \\ \hline
		Avg. Size of embedding File              & 2.43MB-3.78MB & 2.43MB-3.78MB & 270KB             \\ \hline
		Total Number of Model Parameters         & 35,564        & 35,564        & 50,239            \\ \hline
		Avg. Computation Time (for Single Batch) & 45 min        & 45 min        & 3s  \\ \hline
		Embedding Dimensions                     & 8             & 8             & 16                \\ \hline
	\end{tabular}
\end{table}

An end-device's memory and computation power must be taken into account when running a deep learning model. Apart from battery consumption, there are three aspects to consider here: the amount of space the model takes up in your app bundle (number of items to be stored at a time and the space taken by their embedded vectors), the amount of memory it uses at runtime (total number of parameters), and how fast the model runs. \autoref{mem} shows the average on device memory consumption of ReFRS against GMF based FRS. The computation time represents the time taken by the client embedding module for a single batch training. Simulations of our experiments used CPUs with 2.2 GHz processing power, which is equal to any CPU found on most common android devices.  Since ReFRS actively swaps in new interaction sessions with the old ones and does not require a full list of embedded items to update the client model, its memory consumption is far less than any other GMF-based federated model.

\section{Discussion}
In this paper we proposed a federated recommendation system  for end-user device that trains local models on user behavior, aggregates those models in a federated settings while taking into account user's behavioral diversity. There are two main parts to our proposed architecture, a decentralized client module and a semantic server. 
Using a federated server, the ReFRS client model aims to learns the user's interests and interaction behaviors. 
The federated server helps in accelerating the convergence of the client model using semantically aggregated user models.

The ReFRS client model capture users' interaction behavior by developing such a generative model that mimics users' interactions and builds interpretable vector representations.
The goal of the client model is to learn compact vector representations (i.e., embeddings) $\mathbf{z}_s$ of the interaction window $w$ in the latent space, that group semantically similar items together to capture the temporal and contextual relationships between the interacted items.
As a result of this representation, item-sequence relationships should be effectively captured and items that are contextually and temporally similar should be grouped together. 
To avoid storing all item information on the edge device for recommendation, we use a sequential approach where interactions are processed in sessions. These sessions are kept small to fit into limited memory efficiently, and old sessions are swapped out to maximize memory utilization.
As we generate meaningful lightweight vectors, in this work we make use of state-of-the-art recommender models for the next item suggestion.  A major advantage of our client model is that it can effectively handle sparse data and convert it into a size constrained discrete vector representation.

To overcome the non-i.i.d. nature of individual user data, we propose a semantic server that utilizes an efficient grouping scheme that only aggregates semantically similar client models. Rather than grouping clients by requesting and comparing all users' sensitive information like demographic data \cite{muhammad2020fedfast}, our proposed method utilize a neural approach to efficiently capture user affinity by simply looking at the model parameters submitted by each client. We also enable the server to asynchronously update client groupings as the user interaction behavior changes over time. Our sever-side design has three parts, namely a semantic sampler, a clustering unit and a model aggregator, which we will describe below.

We conducted extensive experiments to demonstrate the effectiveness of our proposed method in both federated and centralized settings. In addition, we show the effects of applying Differentially Private Stochastic Gradient Descent and Homomorphic Encryption on the shared parameters and recommend using Homomorphic Encryption to support semantic grouping.

ReFRS solves the cold start, single-user client, and privacy issues with FedFast, but it incurs some security overheads. In particular, the client's model parameter does convey an idea about grouping when viewed with others. Grouping clients with known information will provide insight into other members of the group. 
For future work, we focus on a more web 3.0 friendly SRS i.e. server-less architecture. Rather than being supported by a central server, Web 3.0 apps are built on blockchain, completely decentralized networks of peer-to-peer nodes, or a combination of both. Hence, the client is no longer reliant on a central party to decide on grouping.
In the future, it is vital to explore strategies that can link the proposed decentralized to a server-less architecture.

\section{Conclusion}
We conclude that ReFRS is a promising federated recommendation system that provides user diversity, while preserving user privacy. Unlike other existing methods, ReFRS is highly effective even when dealing with single user-client recommendations.
ReFRS is an FRS designed for edge devices with limited resources that capture diversified user behavior in a decentralized manner. Its main advantages are that it is lightweight, handles data sparsity, captures changes in user behavior, and considers heterogeneity in local models during the global update. This is a more practical paradigm for learning distributed recommender models. 
Due to asynchronicity and neural nature of its federated server, ReFRS is highly efficient and scaleable to accommodate heterogeneous clients. 

\section*{Acknowledgement}
This work is supported by Australian Research Council (Grant No. FT210100624 and DP190101985) and The University of Queensland (Grant No. NS-2103).


\end{document}